\documentclass[twocolumn]{svjour3}          % twocolumn
\smartqed  % flush right qed marks, e.g. at end of proof
\usepackage{graphicx}
\usepackage{amssymb}
\begin{document}

\title{Rapidity Gap Survival in Enhanced Pomeron Scheme}%

\author{Sergey Ostapchenko   \and
   Marcus Bleicher }


\institute{S.\ Ostapchenko \at
Frankfurt Institute for Advanced Studies, 
 60438 Frankfurt am Main, Germany               \\
 D.V. Skobeltsyn Institute of Nuclear Physics,
Moscow State University, 119992 Moscow, Russia\\
              \email{ostapchenko@fias.uni-frankfurt.de}           %  \\
           \and
M.\ Bleicher \at
Frankfurt Institute for Advanced Studies, 
 60438 Frankfurt am Main, Germany               \\   
 Institute for Theoretical Physics, Goethe-Universitat,
 60438 Frankfurt am Main, Germany\\
              \email{bleicher@fias.uni-frankfurt.de} }

\date{Received: date / Accepted: date}

\maketitle

\begin{abstract}
We apply the phenomenological Reggeon field theory framework to investigate
 rapidity gap survival (RGS) probability for diffractive dijet 
production in proton-proton collisions. In particular, we study in some
detail   rapidity gap suppression due to   elastic
 rescatterings of intermediate
partons in the underlying parton cascades, described by enhanced
 (Pomeron-Pomeron interaction) diagrams. We demonstrate that such contributions
 play a subdominant role, compared to the usual, so-called ``eikonal'',
 rapidity gap suppression due to elastic rescatterings of constituent partons
 of the colliding protons. On the other hand, the overall RGS factor
 proves to be sensitive to  color fluctuations in the proton. Hence,
 experimental data on diffractive dijet production can be used to constrain
 the respective model approaches.
\end{abstract}

\section{Introduction}
\label{intro.sec}
An important direction in experimental studies of high energy hadronic
collisions is related to diffractive hadron production, in particular,
to production of high transverse momentum  $p_{\rm t}$  
particles in events characterized 
by large rapidity gaps (RGs) not covered by secondary hadrons. The scientific 
interest to such so-called hard
diffraction phenomena is multifold and related, in particular, to searches
for signatures of new physics in a relatively clean experimental environment
(see Ref.~\cite{kmr17} for a recent review).
On the other hand, the corresponding observables involve both perturbative
and nonperturbative physics and may thus shed some light on the interplay
of the two and  provide an additional insight
into the nonperturbative proton structure.

In contrast to hard diffractive processes  in deep inelastic scattering,
 final states with large rapidity gaps 
 constitute a much smaller fraction of events containing high 
 $p_{\rm t}$ particles  in proton-proton collisions.
 This is because hard processes
typically take place for small values of the impact parameter $b$ between
the colliding protons, where one has a significant overlap  of the 
projectile and target parton clouds, but then, also the probability for 
 additional inelastic rescatterings between protons' constituents is high.
Therefore, there is little chance that a rapidity gap produced in a hard
diffraction process at small $b$
is not covered by secondaries created by the accompanying 
multiple scattering  \cite{dok92}. It has been realized long ago
that the corresponding penalty factor, nicknamed ``rapidity gap survival (RGS)
probability'', results from an interplay between the transverse profile
for a hard diffraction process of interest and the much broader
 inelastic profile for $pp$ collisions  \cite{bjo93}.
 
  Since then, the problem
 has been widely addressed in literature and numerous estimations of the 
 RGS probability for various hard diffraction reactions have been obtained
 \cite{glm93,glm98,glm99,glm06,glm11,glm16,kmr97,kmr01,kmr02,kmr03,kmr09,pet04,fs07,kop07,pas11,lon16,fag17}.
 Most of those studies have  been devoted to the dominant,
 so-called ``eikonal'', mechanism of the
 RG suppression, related to elastic rescatterings between constituent
 partons of the colliding protons, addressing, in particular, the energy
 dependence of the RGS probability \cite{glm98,glm99} and the role of the 
 inelastic diffraction
 treatment in respective models\footnote{Note, however, the arguments of
 Ref.~\cite{fs07} concerning a suppression of contributions of
 inelastic intermediate states.} \cite{glm99,glm06,kmr01,kmr09}.
 Much less understood are the noneikonal  absorptive effects corresponding
 to elastic rescatterings of intermediate partons, for which the obtained
 numerical results differ considerably \cite{glm11,glm16,kmr09}.

In this work, we are going to investigate the RGS probabilities 
 for diffractive dijet production in the framework of the Gribov's
 Reggeon Field Theory (RFT) \cite{gri68}, addressing, in particular,
  in some detail the role of the noneikonal absorption. Our choice was
   partly motivated
 by previous study of soft diffraction by one of us, where such noneikonal
 effects proved to be extremely important, giving rise to huge (up to an
 order of magnitude) corrections to diffractive cross sections \cite{ost10}
 (see, e.g., Fig.~15 in that reference). Since the role of semihard 
 processes, for  relatively small parton transverse momentum, 
 in multiple scattering
 is not too different from the one of purely soft interactions,
 at least in our model, we expected that the noneikonal absorption
 is quite important for diffractive jet production as well.
 
More specifically, we employ the enhanced Pomeron framework
\cite{kan73,kai86,ost06}, as implemented in the QGSJET-II model \cite{ost06a,ost11}.
The approach treats consistently both the usual multiple scattering processes,
describing individual parton cascades as  Pomeron exchanges, and rescatterings
of intermediate  partons in those cascades off the projectile and target
 protons and off each other, which is treated as Pomeron-Pomeron interactions.
Importantly, the latter contributions are resummed to all orders \cite{ost06}.

Hard processes are incorporated in the scheme following the so-called
``semihard Pomeron'' approach \cite{dre01}:
splitting  general parton cascades into soft  and  hard  parts.
 The latter are characterized
by high enough parton virtualities $|q^{2}|>Q_{0}^{2}$, $Q_{0}$
being some cutoff for pQCD being applicable, and are treated by means
of the Dokshitzer-Gribov-Lipatov-Altarelli-Parisi  (DGLAP)\\ evolution equations.
 In turn, the nonperturbative
soft parts involve low-$q^2$ ($|q^{2}|<Q_{0}^{2}$) partons and are described
by  phenomenological soft Pomeron asymptotics.

To treat low mass diffraction and the related absorptive  
effects, a Good-Walker-type \cite{goo60} framework is employed,
considering the interacting protons to be represented by a superposition
of a number of eigenstates which diagonalize the scattering matrix,
characterized by different couplings to Pomerons \cite{kai79}. The
respective partonic interpretation is based on the color fluctuations
picture \cite{fra08}, i.e.~the representation of the proton wave
function by a superposition of parton Fock states of different sizes.
Fock states of larger transverse size are characterized by lower (more
dilute) spatial parton densities,
 while more compact ones are more
densely packed with partons.\footnote{It is noteworthy that the integrated
parton density is, however, lower for  Fock states of smaller size
\cite{kmr01,fra08}.} As will be demonstrated in the following,
such color fluctuations have an important impact on the strength  of the
rapidity gap suppression.

The outline of the paper is as follows. In Section \ref{sec:jet-diffr},
we derive  expressions for cross sections of single and central
diffractive dijet production, introducing step by step the various
absorptive corrections. In Section \ref{sec:results}, we present our
numerical results and discuss them in some detail. Finally, we conclude 
in Section \ref{sec:Outlook}.
\section{Cross sections for diffractive dijet production\label{sec:jet-diffr}}
\begin{figure}
\centering{}
\includegraphics[width=3cm,height=4cm]{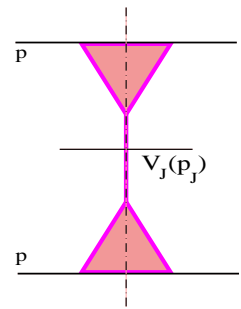}
\caption{Schematic view for 
the general RFT diagram for inclusive jet production
in $pp$ collisions: the projectile and target  ``triangles''  consist
of  fanlike enhanced Pomeron graphs; $V_{J}(p_{J})$ is the parton
$J$ emission vertex from a cut Pomeron. The cut plane is shown by
the vertical dotted-dashed line.\label{fig:2jet}}
\end{figure}

\begin{figure*}
\centering{}
\includegraphics[width=0.9\textwidth,height=2.5cm]{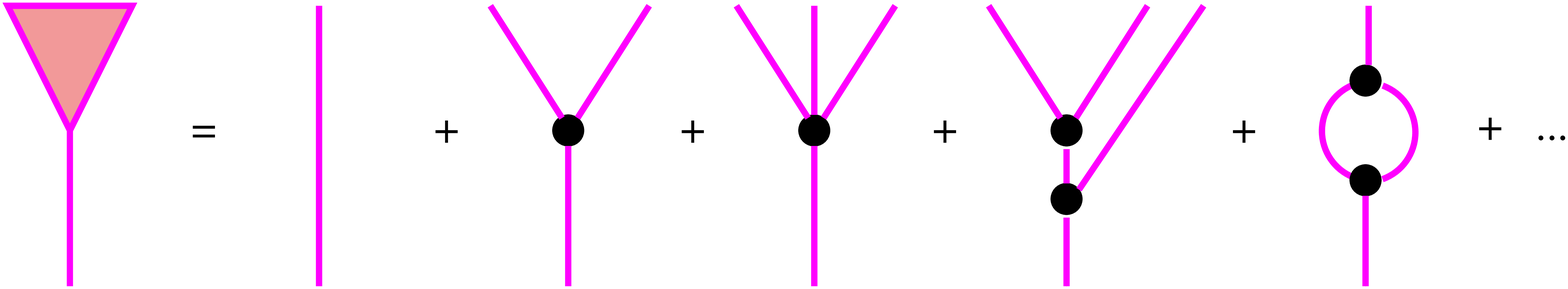}
\caption{Examples of enhanced Pomeron graphs of lowest orders, contributing
to the structure of the projectile and target triangles in Fig.~\ref{fig:2jet};
Pomerons are shown by thick lines and multi-Pomeron vertexes by filled
circles.
\label{fig:triangle}}
\end{figure*}

To set the scene, let us start with the inclusive cross section for 
  high   $p_{{\rm t}}$ jet production. Partial contributions to 
  this cross section from various configurations of proton-proton collisions
  generally involve multiple scattering processes, containing  additional
  soft ($|q^{2}|<Q_{0}^{2}$) and hard ($|q^{2}|>Q_{0}^{2}$) parton cascades
  fragmenting into secondary hadrons, as well as virtual parton cascades
  describing elastic rescatterings between constituent partons of the protons.
  Nevertheless, by virtue of the Abra\-movskii-Gribov-Kancheli (AGK) 
  cancellations \cite{agk}, such multiple scattering processes give zero
  contribution to the inclusive cross section of interest, 
which is described by Kancheli-Mueller-type 
diagrams depicted in Fig.~\ref{fig:2jet}. %
 The internal structure of the projectile and target  triangles
in Fig.~\ref{fig:2jet}  is explained in Fig.~\ref{fig:triangle}:
it contains both the  basic contribution of  an ``elementary'' parton cascade
described as a single Pomeron emission by the parent
hadron {[}1st graph in the right-hand side (rhs)  of  Fig.~\ref{fig:triangle}{]}
  and various absorptive corrections to that process 
due to  rescatterings of intermediate partons in the cascade off
the parent hadron  and off each other. %

As a result, we obtain the usual collinear factorization ansatz for  
 the inclusive cross section
 $\sigma_{pp}^{2{\rm jet}}(s,p_{{\rm t}}^{{\rm cut}})$
for the production of a pair of jets of transverse momentum
 $p_{{\rm t}}>p_{{\rm t}}^{{\rm cut}}$:
\begin{eqnarray}
\sigma_{pp}^{2{\rm jet}}(s,p_{{\rm t}}^{{\rm cut}})= \int \! d^2b \, d^2b'
\int\! dx^{+}\, dx^{-}\int_{p_{{\rm t}}>p_{{\rm t}}^{{\rm cut}}}\! dp_{{\rm
t}}^{2} &&
\nonumber \\\times \;  \sum_{I,J=q,\bar{q},g}
\frac{d\sigma_{IJ}^{2\rightarrow2}(x^{+}x^{-}s,p_{{\rm t}}^{2})}
{dp_{{\rm t}}^{2}}\;
G_{I}(x^{+},M_{{\rm F}}^{2},b')&&
\nonumber \\\times \;
 G_{J}(x^{-},M_{{\rm F}}^{2},|\vec{b}-\vec{b}'|)
 \,, \label{eq:sig-2jet} &&
\end{eqnarray}
where for future convenience we keep the impact parameter $b$ dependence
and express the  integrand in the rhs of Eq.~(\ref{eq:sig-2jet}) via
 the generalized
parton distributions (GPDs) in impact parameter space 
 $G_{I}(x,Q^{2},b)$, instead of the usual integrated parton distribution
 functions (PDFs), 
 $f_{I}(x,Q^{2})=\int  \! d^2b\;G_{I}(x,Q^{2},b)$. Here $s$ is the center
 of mass (c.m.) energy squared, 
$x^{\pm}$ -  parton light-cone momentum fractions,
 $M_{{\rm F}}^{2}$ - the factorization scale, and 
$d\sigma_{IJ}^{2\rightarrow2}(\hat s,p_{t}^{2})/dp_{{\rm t}}^{2}$ is the
parton scatter cross section.

\begin{figure}
\centering{}
\includegraphics[width=0.45\textwidth,height=3cm]{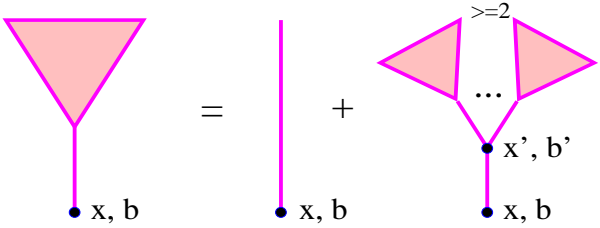}
\caption{Recursive equation for a fan diagram contribution
 $\chi_{(i)}^{{\rm fan}}(\hat{s},b)$,
$\hat{s}=s_{0}/x$.
\label{fig:ffan}}
\end{figure}

The GPDs for arbitrary  $Q^{2}>Q_{0}^{2}$ are obtained  evolving
the input  ones from the cutoff scale $Q_{0}^{2}$:
\begin{eqnarray}
G_{I}(x,Q^{2},b) =\sum_{I'}\int_{x}^{1}\!\frac{dz}{z}\, 
E_{I'\rightarrow I}(z,Q_{0}^{2},Q^{2})&& \nonumber \\
\times \; G_{I'}(x/z,Q_{0}^{2},b)\,,
\label{eq:GPD-evolv}
\end{eqnarray}
with $E_{I\rightarrow J}(z,q^{2},Q^{2})$ being the solution
of the DGLAP equations for the initial condition 
 $E_{I\rightarrow J}(z,q^{2},q^{2})=\delta_{I}^{J}\,\delta(1-z)$.
 In turn, $G_{I}(x,Q_0^{2},b)$ is defined 
 summing over   partial contributions of  different diffractive
 eigenstates  $|i\rangle$ of the proton, with the partial weights $C_{i}$,
  as\footnote{To simplify the discussion, we neglect
 here  Pomeron ``loop'' contributions, exemplified by the last graph 
 in the rhs of Fig.~\ref{fig:triangle}, the complete treatment
 being described in Refs.~\cite{ost10,ost11}.} \cite{ost06a,ost16}
\begin{eqnarray}
x\,G_{I}(x,Q_0^{2},b) =\sum_i C_i \left\{\chi_{(i)I}^{\mathbb{P}}(s_{0}/x,b)
 \right. \nonumber &&\\
+ \;G\int\! d^{2}b'\int\!\frac{dx'}{x'} 
 \; \chi_{\mathbb{P}I}^{\mathbb{P}}(s_{0}\,x'/x,|\vec{b}-\vec{b}'|) \nonumber &&\\
\times  \left.\left[1-e^{-\chi_{(i)}^{{\rm fan}}(s_{0}/x',b')}
-\chi_{(i)}^{{\rm fan}}(s_{0}/x',b')\right] 
\right\} \!,\label{eq:GPD-Q0} &&
\end{eqnarray}
being expressed via the solution $\chi_{(i)}^{{\rm fan}}$  of the ``fan''
 diagram equation of Fig.~\ref{fig:ffan},
\begin{eqnarray}
\chi_{(i)}^{{\rm fan}}(\hat{s},b)=
\chi_{(i)\mathbb{P}}^{\mathbb{P}}(\hat{s},b)\nonumber &&\\
+\;G\int\! d^{2}b'\int\!\frac{dx'}{x'}\;
 \chi_{\mathbb{PP}}^{\mathbb{P}}(x'\hat{s},|\vec{b}-\vec{b}'|) \nonumber &&\\
\times \left[1-e^{-\chi_{(i)}^{{\rm fan}}(s_{0}/x',b')}
-\chi_{(i)}^{{\rm fan}}(s_{0}/x',b')\right] \!.&&
\label{eq:fan}
\end{eqnarray}
In Eqs.~(\ref{eq:GPD-Q0}-\ref{eq:fan}), $s_0=1$ GeV$^2$ is the hadronic mass
scale and the eikonals 
$\chi_{(i)\mathbb{P}}^{\mathbb{P}}$ and
$\chi_{\mathbb{PP}}^{\mathbb{P}}$
correspond  to   Pomeron exchanges between the proton  diffractive
eigenstate $|i\rangle$ and a multi-Pomeron vertex or, respectively,
 between two multi-Pomeron vertexes, while
 $\chi_{(i)I}^{\mathbb{P}}$ and 
  $\chi_{\mathbb{P}I}^{\mathbb{P}}$ describe  Pomerons
  coupled to   parton $I$ on one side and to the proton 
  represented by its diffractive eigenstate $|i\rangle$ or, respectively,
  to a multi-Pomeron vertex, on the other side, as discussed in more detail in
  \cite{ost06a,ost11}. It is easy to see that the expression in the curly
  brackets in Eq.~(\ref{eq:GPD-Q0}) is obtained from the rhs of Eq.~(\ref{eq:fan})
under the replacements $\chi_{(i)\mathbb{P}}^{\mathbb{P}}
\rightarrow \chi_{(i)I}^{\mathbb{P}}$,
 $\chi_{\mathbb{PP}}^{\mathbb{P}} \rightarrow 
 \chi_{\mathbb{P}I}^{\mathbb{P}}$, 
 i.e.\ by picking up parton $I$ from the downmost Pomeron.

It is noteworthy that Eqs.~(\ref{eq:GPD-evolv}-\ref{eq:fan})
have been derived in Ref.\  \cite{ost06a},  neglecting
parton transverse diffusion during the perturbative ($|q^{2}|>Q_{0}^{2}$)
evolution and assuming Pomeron-Pomeron interactions to be mediated by
nonperturbative parton processes, using the vertexes for the transition
of $m$ into $n$ Pomerons  of the form  \cite{kai86}:
$G^{(m,n)}=G\,\gamma_{\mathbb{P}}^{m+n}$, where
 $G$ is related to the triple-Pomeron coupling $r_{3\mathbb{P}}$
as $G=r_{3\mathbb{P}}/(4\pi\gamma_{\mathbb{P}}^{3}$).
For smaller $x$, the soft  ($|q^{2}|<Q_{0}^{2}$) parton
evolution proceeds over a longer rapidity interval and   results in a larger
transverse spread of the parton cloud at the scale $Q_{0}^{2}$, due to the
 transverse diffusion. On the other hand,
 for a higher scale  $Q^{2}$, a larger part of the available rapidity range
 is ``eaten'' by the perturbative evolution [c.f.~Eq.~(\ref{eq:GPD-evolv})].
  As a consequence, for
a given $x$, partons of higher $Q^{2}$ are distributed over a smaller
transverse area.%
\begin{figure*}
\centering{}
\includegraphics[width=0.65\textwidth,height=3.cm]{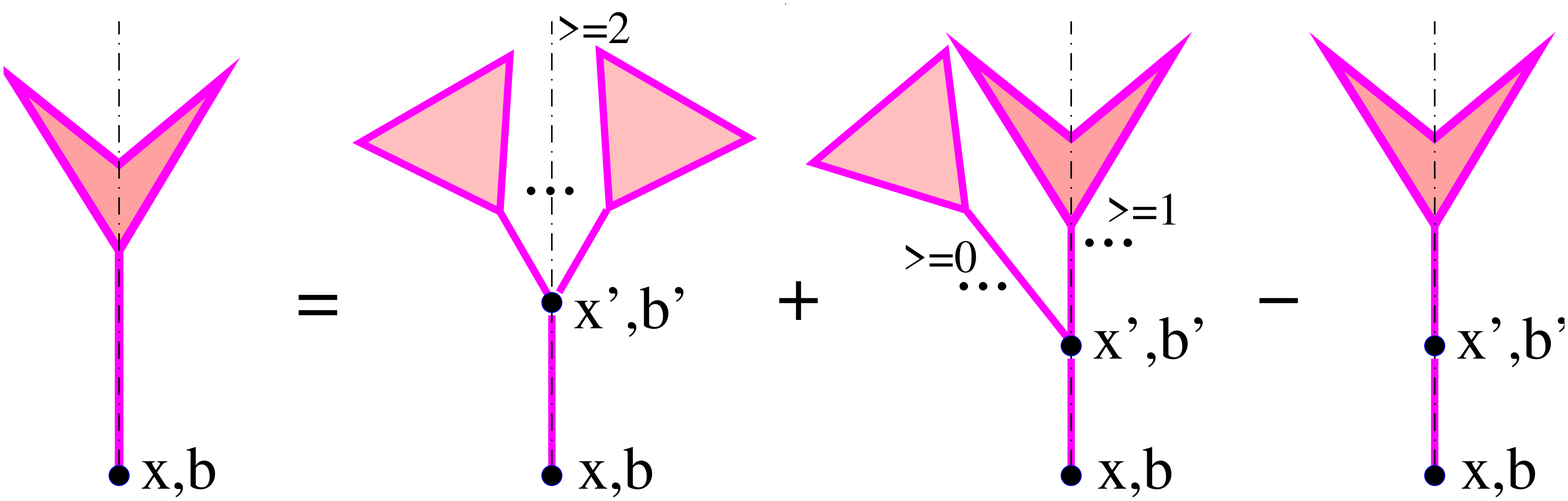}
\caption{Recursive equation for the  contribution 
$2\chi_{(i)}^{{\rm fan(D)}}(\hat{s},b,y^{\rm gap})$ of diffractive
cuts of fan diagrams of  Fig.~\ref{fig:ffan}, $\hat{s}=s_{0}/x$. 
The cut plane is shown by the vertical dotted-dashed lines.
\label{fig:ffan-D}}
\end{figure*}%

If we naively assumed the same kind of factorization for diffractive dijet
production, the respective cross sections would be defined by  subsets of
cut diagrams corresponding to Fig.~\ref{fig:2jet}, characterized by a
desirable structure of rapidity gaps. For example, for the case of central
hard diffraction (``double Pomeron exchange''), with the forward and backward
rapidity gaps being larger than $y^{\rm gap}$, we would  obtain
\begin{eqnarray}
\sigma_{pp}^{2{\rm jet-DPE(fact)}}(s,p_{{\rm t}}^{{\rm cut}}
,y^{\rm gap})= \int \! d^2b \, d^2b'
\int\! dx^{+}dx^{-} &&
\nonumber \\
\times  \int_{p_{{\rm t}}>p_{{\rm t}}^{{\rm cut}}}\! dp_{{\rm
t}}^{2} \;
\sum_{I,J=q,\bar{q},g} 
\frac{d\sigma_{IJ}^{2\rightarrow2}(x^{+}x^{-}s,p_{{\rm t}}^{2})}
{dp_{{\rm t}}^{2}}&&
\nonumber \\
\times \;G^D_{I}(x^{+},M_{{\rm F}}^{2},b',y^{\rm gap})\,
 G^D_{J}(x^{-},M_{{\rm F}}^{2},|\vec{b}-\vec{b}'|,y^{\rm gap})
 \,. \label{eq:sig-2jet-dpe-fact}&&
\end{eqnarray}
Here the diffractive GPDs  $G^D_{I}(x,Q^{2},b,y^{\rm gap})$ for an
arbitrary scale  $Q^{2}$ are obtained  via DGLAP evolution from  $Q_{0}^{2}$
till  $Q^{2}$ [similarly to Eq.~(\ref{eq:GPD-evolv})], while\\ 
$G^D_{I}(x,Q_0^{2},b,y^{\rm gap})$ is expressed via the contribution 
 $2\chi_{(i)}^{{\rm fan(D)}}$ of diffractive cuts of the fan diagrams
  of  Fig.~\ref{fig:ffan} as
\begin{eqnarray}
x\,G^D_{I}(x,Q_0^{2},b,y^{\rm gap})= \sum_i C_i
\left\{\frac{G}{2}\int\! d^{2}b'\int\!\frac{dx'}{x'}   \right. && \nonumber \\
\times \; \Theta (-\ln x'-y^{\rm gap}) \;
 \chi_{\mathbb{P}I}^{\mathbb{P}}(s_{0}\, x'/x,|\vec{b}-\vec{b}'|)
  && \nonumber \\
\times \left[(1-e^{-\chi_{(i)}^{{\rm fan}}(s_{0}/x',b')})^2
+ (e^{2\chi_{(i)}^{{\rm fan(D)}}(s_{0}/x',b',y^{\rm gap})}-1)
 \right.  && \nonumber \\
\times  \left.\left. e^{-2\chi_{(i)}^{{\rm fan}}(s_{0}/x',b')} 
- 2\chi_{(i)}^{{\rm fan(D)}}(s_{0}/x',b',y^{\rm gap}) \right]
 \right\}\!.\label{eq:GPD-D}&&
\end{eqnarray}
The latter are defined by the recursive equation of  Fig.~\ref{fig:ffan-D}:
\begin{eqnarray}
2\chi_{(i)}^{{\rm fan(D)}}(\hat s,b,y^{\rm gap})=
G\int\! d^{2}b'\int\!\frac{dx'}{x'}  && \nonumber \\
\times \; \Theta (-\ln x'-y^{\rm gap})\;
  \chi_{\mathbb{PP}}^{\mathbb{P}}(x'\hat s,|\vec{b}-\vec{b}'|)
  && \nonumber \\
\times  \left[(1-e^{-\chi_{(i)}^{{\rm fan}}(s_{0}/x',b')})^2
+ (e^{2\chi_{(i)}^{{\rm fan(D)}}(s_{0}/x',b',y^{\rm gap})}-1) \right.
  && \nonumber \\
\times \left. e^{-2\chi_{(i)}^{{\rm fan}}(s_{0}/x',b')} \;
-2\chi_{(i)}^{{\rm fan(D)}}(s_{0}/x',b',y^{\rm gap}) \right]\!.
 &&\label{eq:ffan-D}
\end{eqnarray}
 Similarly to  Eqs.~(\ref{eq:GPD-Q0}-\ref{eq:fan}), the expression in the curly
  brackets in Eq.~(\ref{eq:GPD-D}) is obtained from the rhs of 
   Eq.~(\ref{eq:ffan-D}) under the replacement 
 $\chi_{\mathbb{PP}}^{\mathbb{P}} \rightarrow 
 \chi_{\mathbb{P}I}^{\mathbb{P}}$. 

The analog of  Eq.~(\ref{eq:sig-2jet-dpe-fact}) for single (here, projectile)
hard diffraction is\footnote{Strictly speaking, Eq.~(\ref{eq:sig-2jet-sd-fact})
contains also the contribution of double diffraction, corresponding to
a dissociation of the projectile proton into a low mass hadronic system,
in addition to the formation of a high mass state on the target side
(see the discussion in Refs.~\cite{ost10,ost14}). Similarly, 
Eq.~(\ref{eq:sig-2jet-dpe-fact}) accounts also for situations when
the projectile or/and target protons are excited into low mass hadronic states,
in addition to the formation of the central diffractive system.}
\begin{eqnarray}
\sigma_{pp}^{2{\rm jet-SD(fact)}}(s,p_{{\rm t}}^{{\rm cut}}
,y^{\rm gap}) = \int \! d^2b \, d^2b'\int\! dx^{+}dx^{-}
\nonumber &&\\
\times 
\int_{p_{{\rm t}}>p_{{\rm t}}^{{\rm cut}}}\! dp_{{\rm t}}^{2}
 \; \sum_{I,J=q,\bar{q},g}
\frac{d\sigma_{IJ}^{2\rightarrow2}(x^{+}x^{-}s,p_{{\rm t}}^{2})}
{dp_{{\rm t}}^{2}}\nonumber &&\\
\times  \;G^D_{I}(x^{+},M_{{\rm F}}^{2},b',y^{\rm gap})\,
 G_{J}(x^{-},M_{{\rm F}}^{2},|\vec{b}-\vec{b}'|)
 \,. \label{eq:sig-2jet-sd-fact} &&
\end{eqnarray}

\begin{figure}
\centering{}
\includegraphics[width=0.4\textwidth,height=4.5cm]{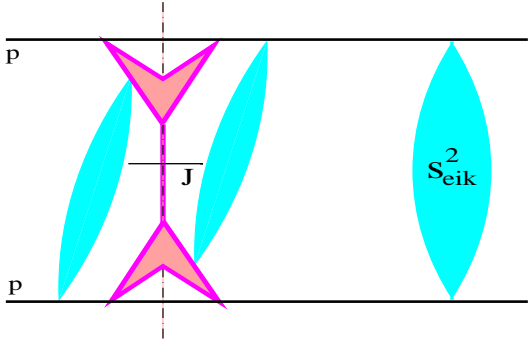}
\caption{Schematic view for 
 central hard
diffraction;  parton $J$ is emitted from a cut Pomeron at the central rapidity,
the cut plane  shown by the vertical dotted-dashed line. 
Eikonal absorption due to constituent parton rescatterings
is shown symbolically by the vertical ellipse marked ``$S^2_{\rm eik}$'';
noneikonal absorptive corrections due to  rescatterings of intermediate 
partons mediating the diffractive scattering are indicated by inclined 
ellipses.
\label{fig:ddijets}}
\end{figure}

Since the integrated diffractive PDFs 
$f^D_{I}(x,Q^{2},y^{\rm gap})=  \int \! d^2b \,
G^D_{I}(x,Q^{2},b,y^{\rm gap})$ can be inferred from experimental
studies of diffractive deep inelastic scattering, 
Eqs.~(\ref{eq:sig-2jet-dpe-fact}) and (\ref{eq:sig-2jet-sd-fact}) could have
been
well-defined predictions. In reality, there is no good reason to assume such
kind of factorization for not fully inclusive quantities, like diffractive
cross sections, and the real picture is significantly more complicated,
as shown symbolically in Fig.~\ref{fig:ddijets}. %

First, the expressions in the rhs of 
Eqs.~(\ref{eq:sig-2jet-dpe-fact}) and  (\ref{eq:sig-2jet-sd-fact}) have to be
supplemented by the probability that the desirable rapidity gaps are not
filled by secondary particles produced in additional inelastic scatterings
processes between constituent partons of the projectile and target protons.
For given diffractive eigenstates  $|i\rangle$,  $|j\rangle$ of the two
 protons and impact parameter $b$, the corresponding RGS probability is 
 $\exp (-\Omega _{ij}(s,b))$, where the so-called opacity $\Omega _{ij}$
 is defined as twice the sum over  imaginary parts of all significant
 irreducible Pomeron graphs coupled to the eigenstates  $|i\rangle$
 and  $|j\rangle$. 
 
 A relatively compact expression for  $\Omega _{ij}$ has been obtained in
 \cite{ost06} summing the contributions of arbitrary Pomeron ``nets''
 exchanged between the projectile and target protons:\footnote{As mentioned
 above, in our discussion we neglect for simplicity the contributions of
 graphs containing Pomeron loops. In numerical calculations, presented in 
 Section~\ref{sec:results}, we use the complete formalism of the 
 QGSJET-II model, Pomeron loop contributions included.}
\begin{eqnarray}
\Omega _{ij}(s,b)=2\chi_{ij}^{\mathbb{P}}(s,b)+2G\int\! d^{2}b'\int\!\frac{dx'}{x'}
\left\{(1\right.
 &&\nonumber \\
-\;e^{-\chi_{(i)|(j)}^{{\rm net}}(s_{0}/x',\vec{b}'|s,\vec{b})})\,
(1-e^{-\chi_{(j)|(i)}^{{\rm net}}(x' s,\vec{b}-\vec{b}'|s,\vec{b})})
 &&\nonumber \\
-\;\chi_{(i)|(j)}^{{\rm net}}(s_{0}/x',\vec{b}'|s,\vec{b}) \,
 \chi_{(j)|(i)}^{{\rm net}}(x' s,\vec{b}-\vec{b}'|s,\vec{b})
 &&\nonumber \\
-\, (\chi_{(i)|(j)}^{{\rm net}}(s_{0}/x',\vec{b}'|s,\vec{b})
-\chi_{(i)\mathbb{P}}^{\mathbb{P}}(s_{0}/x',b'))  &&\nonumber \\
\times \left[(1-e^{-\chi_{(j)|(i)}^{{\rm net}}(x'
s,\vec{b}-\vec{b}'|s,\vec{b})})\,
e^{-\chi_{(i)|(j)}^{{\rm net}}(s_{0}/x',\vec{b}'|s,\vec{b})}
\right. &&\nonumber \\
- \left.\left.
\chi_{(j)|(i)}^{{\rm net}}(x' s,\vec{b}-\vec{b}'|s,\vec{b})\right]\right\}\!, &&
\label{chi-tot}
\end{eqnarray}
where $\chi_{ij}^{\mathbb{P}}(s,b)$ is the eikonal for a single Pomeron exchange
between the eigenstates  $|i\rangle$ and  $|j\rangle$ 
while the ``net-fan'' eikonal
$\chi_{(i)|(j)}^{{\rm net}}$ corresponds to the summary contribution of arbitrary
irreducible Pomeron nets exchanged between  the projectile and target protons
(represented by the  eigenstates  $|i\rangle$ and  $|j\rangle$)
and coupled to a given multi-Pomeron vertex, which is defined by the recursive
equation [c.f.~Eq.~(\ref{eq:fan})]:
\begin{eqnarray}
\chi_{(i)|(j)}^{{\rm net}}(\hat s,\vec{b}''|s,\vec{b})=
\chi_{(i)\mathbb{P}}^{\mathbb{P}}(\hat{s},b'')
+G\int\! d^{2}b'\int\!\frac{dx'}{x'}
&&\nonumber \\
\times \; \chi_{\mathbb{PP}}^{\mathbb{P}}(x'\hat{s},|\vec{b}''-\vec{b}'|)
\left[(1-e^{-\chi_{(i)|(j)}^{{\rm net}}(s_{0}/x',\vec{b}'|s,\vec{b})})\right.
&&\nonumber \\
\times \left. e^{-\chi_{(j)|(i)}^{{\rm net}}(x's,\vec{b}-\vec{b}'|s,\vec{b})}
 -\chi_{(i)|(j)}^{{\rm net}}(s_{0}/x',\vec{b}'|s,\vec{b})\right]\!.&&
 \label{net-fan}\end{eqnarray}

Taking into consideration only the above-discussed  eikonal
rapidity gap suppression, shown symbolically by the vertical ellipse in 
Fig.~\ref{fig:ddijets},
Eqs.~(\ref{eq:sig-2jet-dpe-fact}) and  (\ref{eq:sig-2jet-sd-fact}) will change to
\begin{eqnarray}
\sigma_{pp}^{2{\rm jet-DPE(eik)}}(s,p_{{\rm t}}^{{\rm cut}}
,y^{\rm gap}) =   \int \! d^2b \, d^2b'
\int\! dx^{+}dx^{-}
\nonumber && \\
\times \int_{p_{{\rm t}}>p_{{\rm t}}^{{\rm cut}}}\! dp_{{\rm t}}^{2}\; 
\sum_{I,J=q,\bar{q},g}
\frac{d\sigma_{IJ}^{2\rightarrow2}(x^{+}x^{-}s,p_{{\rm t}}^{2})}
{dp_{{\rm t}}^{2}}
\nonumber && \\
\times \; \sum_{i,j} C_i\, C_j \;
G^D_{I(i)}(x^{+},M_{{\rm F}}^{2},b',y^{\rm gap})
\nonumber && \\
\times \; G^D_{J(j)}(x^{-},M_{{\rm F}}^{2},|\vec{b}-\vec{b}'|,y^{\rm gap})\;
 e^{-\Omega _{ij}(s,b)}
 \label{eq:sig-2jet-dpe-eik} && \\
\sigma_{pp}^{2{\rm jet-SD(eik)}}(s,p_{{\rm t}}^{{\rm cut}}
,y^{\rm gap}) = \int \! d^2b \, d^2b'
\int\! dx^{+}dx^{-}
\nonumber && \\
\times \int_{p_{{\rm t}}>p_{{\rm t}}^{{\rm cut}}}\! dp_{{\rm t}}^{2}\;
 \sum_{I,J=q,\bar{q},g}
\frac{d\sigma_{IJ}^{2\rightarrow2}(x^{+}x^{-}s,p_{{\rm t}}^{2})}
{dp_{{\rm t}}^{2}}
\nonumber && \\
\times \;  \sum_{i,j} C_i\, C_j \;
G^D_{I(i)}(x^{+},M_{{\rm F}}^{2},b',y^{\rm gap}) 
\nonumber && \\
\times \; G_{J(j)}(x^{-},M_{{\rm F}}^{2},|\vec{b}-\vec{b}'|)\;
 e^{-\Omega _{ij}(s,b)}
 , \label{eq:sig-2jet-sd-eik} &&
\end{eqnarray}
where $G_{I(i)}$ and $G^D_{I(i)}$ are obtained evolving from $Q_0^2$ till
$M_{{\rm F}}^{2}$ the corresponding
partial contributions [expressions in the curly brackets in 
Eqs.~(\ref{eq:GPD-Q0}) and (\ref{eq:GPD-D}), respectively] of the eigenstate
  $|i\rangle$.
  
   Neglecting color fluctuations in the interacting protons, i.e.~considering
   a single eigenstate $i\equiv 1$, would significantly simplify the analysis
   since the total opacity $\Omega _{pp}(s,b)$ can be inferred from 
   measurements of the differential elastic $pp$ cross section.
   Yet, as  already stressed previously \cite{bjo93,kmr01,kmr03},
    even in such a case  the overall RGS factor would not be 
 a    universal constant,  depending generally  on the 
   process under study and the respective kinematics. In the particular case
   of diffractive dijet production, considered here,  a higher
   jet transverse momentum cutoff $p_{{\rm t}}^{{\rm cut}}$ implies 
   a lower probability for the rapidity gap survival, since
   a larger part
   of the available rapidity range will be ``eaten'' by the DGLAP evolution
   of  $G_{I(i)}$ and $G^D_{I(i)}$
   in the high $q^2$ range [c.f.~Eq.~(\ref{eq:GPD-evolv})]. Hence, a smaller
   part will be left for parton transverse diffusion during the
    soft evolution  at $|q^{2}|<Q_{0}^{2}$, with the end result that
    the contribution of moderately large impact parameters $b$ to the 
    integrands of  Eqs.~(\ref{eq:sig-2jet-dpe-eik}-\ref{eq:sig-2jet-sd-eik})
    will be reduced. On the other hand, diffractive production at small $b$
     is strongly suppressed by a higher opacity
     $\Omega _{pp}(s,b)$, which reflects a higher probability for additional
     inelastic scattering processes, 
     due to a stronger overlap of  parton clouds
     of the interacting protons.
     
     While Eqs.~(\ref{eq:GPD-Q0}) and (\ref{eq:GPD-D}) already account for 
     absorptive corrections to  $G_{I}$ and $G^D_{I}$ due to 
     rescatterings of intermediate partons off their parent protons,
     additional suppression, shown symbolically by the inclined ellipses
     in Fig.~\ref{fig:ddijets}, comes from their elastic rescatterings off 
     the partner protons. Intermediate partons in the cascades mediating those 
   rescatterings may in turn scatter elastically off the initial protons, etc.
     Taking these effects into consideration, we obtain, similarly to the
     case of soft diffraction  in Refs.~\cite{ost10,ost06a},
      the cross sections  for central  and single (here, projectile) 
   diffractive dijet production  as % 
\begin{figure*}
\centering{}
\includegraphics[width=0.65\textwidth,height=3.5cm]{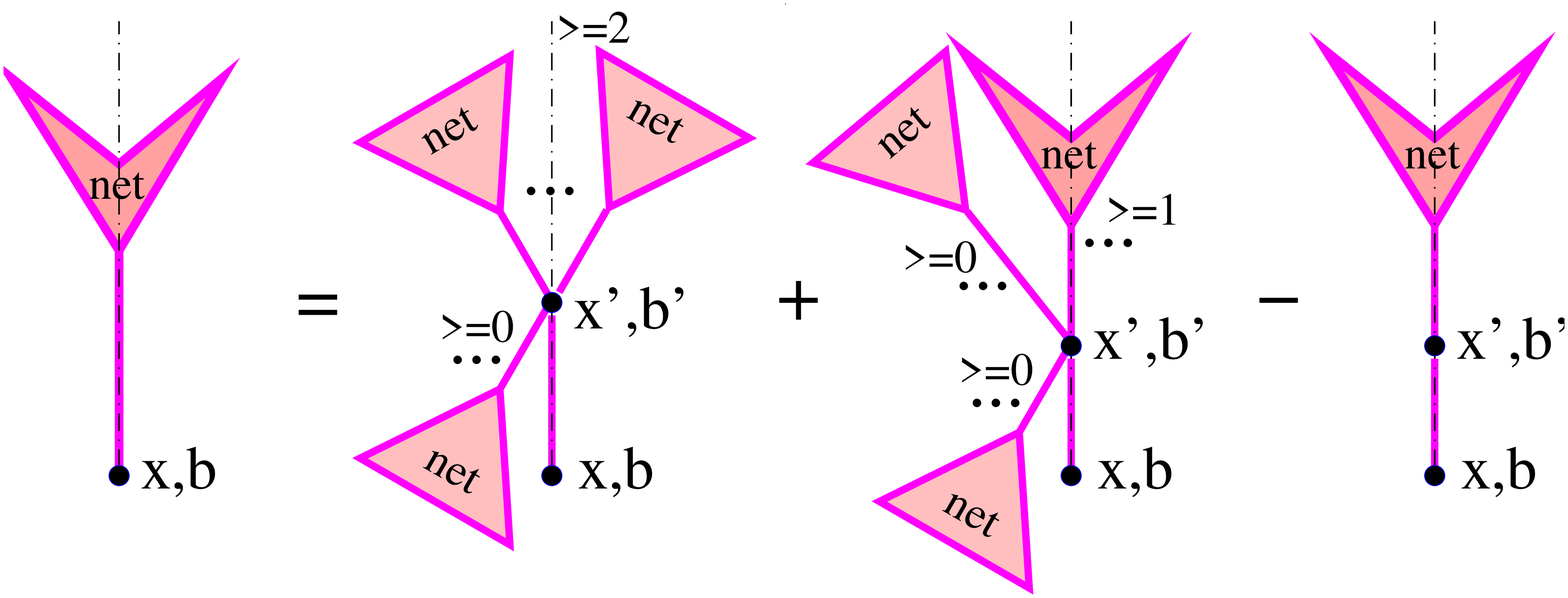}
\caption{Recursive equation for the  contribution 
$2\chi_{(i)|(j)}^{{\rm net(D)}}$ of diffractive
cuts of net-fan diagrams; the cut plane is shown by
the vertical dotted-dashed lines.
\label{fig:ffan-D'}}
\end{figure*}% 
\begin{eqnarray}
\sigma_{pp}^{2{\rm jet-DPE}}(s,p_{{\rm t}}^{{\rm cut}}
,y^{\rm gap}) = \int \! d^2b \, d^2b'
\int\! dx^{+}dx^{-}
\nonumber  && \\
\times  \int_{p_{{\rm t}}>p_{{\rm t}}^{{\rm cut}}}\! dp_{{\rm
t}}^{2}\; \sum_{I,J=q,\bar{q},g}
\frac{d\sigma_{IJ}^{2\rightarrow2}(x^{+}x^{-}s,p_{{\rm t}}^{2})}
{dp_{{\rm t}}^{2}}
\nonumber  && \\
\times   \sum_i C_i\, C_j\;
 \tilde G^D_{I(i)|(j)}(x^{+},M_{{\rm F}}^{2},\vec b',y^{\rm gap}|s,\vec b)
\nonumber  && \\
\times \; \tilde G^D_{J(j)|(i)}(x^{-},M_{{\rm F}}^{2},\vec{b}-\vec{b}',y^{\rm gap}|s,\vec b)\;
 e^{-\Omega _{ij}(s,b)}
 \label{eq:sig-2jet-dpe} && \\
\sigma_{pp}^{2{\rm jet-SD}}(s,p_{{\rm t}}^{{\rm cut}}
,y^{\rm gap}) = \int \! d^2b \, d^2b'
\int\! dx^{+}dx^{-} 
\nonumber  && \\
\times \int_{p_{{\rm t}}>p_{{\rm t}}^{{\rm cut}}}\! dp_{{\rm
t}}^{2} \;\sum_{I,J=q,\bar{q},g}
\frac{d\sigma_{IJ}^{2\rightarrow2}(x^{+}x^{-}s,p_{{\rm t}}^{2})}
{dp_{{\rm t}}^{2}}
\nonumber && \\
\times   \sum_i C_i\, C_j \;
 \tilde G^D_{I(i)|(j)}(x^{+},M_{{\rm F}}^{2},\vec b',y^{\rm gap}|s,\vec b)
\nonumber  && \\
\times \; \tilde  G_{J(j)|(i)}(x^{-},M_{{\rm F}}^{2},\vec{b}-\vec{b}'|s,\vec b)\;
 e^{-\Omega _{ij}(s,b)}
, \label{eq:sig-2jet-sd} &&
\end{eqnarray}
where  $\tilde G_{I(i)|(j)}$ and $\tilde G^D_{I(i)|(j)}$ now depend explicitly
on the geometry of the $pp$ collision, being defined at the 
$Q_0^{2}$-scale  as
\begin{eqnarray}
x\,\tilde G_{I(i)|(j)}(x,Q_0^{2},\vec b''|s,\vec b) =
\chi_{(i)I}^{\mathbb{P}}(s_{0}/x,b'') \nonumber &&\\
+\;G\int\! d^{2}b'\int\!\frac{dx'}{x'} \;
\chi_{\mathbb{P}I}^{\mathbb{P}}(s_{0}\,x'/x,|\vec{b''}-\vec{b}'|)
  \nonumber &&\\
\times 
\left\{(1-e^{-\chi_{(i)|(j)}^{{\rm net}}(s_{0}/x',\vec{b}'|s,\vec{b})})\,
 e^{-2\chi_{(j)|(i)}^{{\rm net}}(x's,\vec{b}-\vec{b}'|s,\vec{b})}\right.
   \nonumber &&\\
 -\left. \chi_{(i)|(j)}^{{\rm net}}(s_{0}/x',\vec{b}'|s,\vec{b})\right\}
 \label{eq:GPD'-Q0} && \\
x\,\tilde G^D_{I(i)|(j)}(x,Q_0^{2},\vec b'',y^{\rm gap}|s,\vec b) =
\frac{G}{2}\int\! d^{2}b'\int\!\frac{dx'}{x'}
 \nonumber &&\\
\times \; \Theta (-\ln x'-y^{\rm gap})\;
 \chi_{\mathbb{P}I}^{\mathbb{P}}(s_{0}\, x'/x,|\vec{b}''-\vec{b}'|)
   \nonumber &&\\
\times \left\{(1-
e^{-\chi_{(i)|(j)}^{{\rm net}}(s_{0}/x',\vec{b}'|s,\vec{b})})^2
\right.
  \nonumber &&\\
\times \; e^{-2\chi_{(j)|(i)}^{{\rm net}}(x's,\vec{b}-\vec{b}'|s,\vec{b})}
+ (e^{2\chi_{(i)|(j)}^{{\rm net(D)}}(s_{0}/x',\vec b',y^{\rm gap}|s,\vec{b})}
  \nonumber &&\\
  -\;1) 
\; e^{-2\chi_{(i)|(j)}^{{\rm net}}\!(s_{0}/x',\vec{b}'|s,\vec{b})
-2\chi_{(j)|(i)}^{{\rm net}}(x's,\vec{b}-\vec{b}'|s,\vec{b})}
 \nonumber &&\\
-\left. 2\chi_{(i)|(j)}^{{\rm net(D)}}(s_{0}/x',\vec b',y^{\rm
gap}|s,\vec{b})\right\}\!.  \label{eq:GPD'-D} &&
\end{eqnarray}   
    Here the total contribution
  $2\chi_{(i)|(j)}^{{\rm net(D)}}$     of 
 all the unitarity cuts of the net-fan diagrams, characterized by the 
 desirable rapidity gap signature, is defined by the recursive equation of  
 Fig.~\ref{fig:ffan-D'}
[c.f.~Fig.~\ref{fig:ffan-D} and Eq.~(\ref{eq:ffan-D})]:
\begin{eqnarray}
2\chi_{(i)|(j)}^{{\rm net(D)}}(\hat s,\vec b'',y^{\rm gap}|s,\vec b) =
G\int\! d^{2}b'\int\!\frac{dx'}{x'}
  \nonumber &&\\
\times \; \Theta (-\ln x'-y^{\rm gap})\;
 \chi_{\mathbb{PP}}^{\mathbb{P}}(x'\hat s,|\vec{b}''-\vec{b}'|)
   \nonumber &&\\
\times \left\{(1-e^{-\chi_{(i)|(j)}^{{\rm net}}(s_{0}/x',\vec{b}'|s,\vec{b})})^2
 e^{-2\chi_{(j)|(i)}^{{\rm net}}(x's,\vec{b}-\vec{b}'|s,\vec{b})}\right.
  \nonumber &&\\
+(e^{2\chi_{(i)|(j)}^{{\rm net(D)}}(s_{0}/x',\vec b',y^{\rm gap}|s,\vec{b})}
-1)   \nonumber &&\\
\times \; e^{-2\chi_{(i)|(j)}^{{\rm net}}\!(s_{0}/x',\vec{b}'|s,\vec{b})
-2\chi_{(j)|(i)}^{{\rm net}}(x's,\vec{b}-\vec{b}'|s,\vec{b})}
  \nonumber &&\\
-\left. 2\chi_{(i)|(j)}^{{\rm net(D)}}(s_{0}/x',\vec b',y^{\rm
gap}|s,\vec{b})\right\}\!.  \label{eq:ffan-D'} &&
\end{eqnarray}   
Clearly, the rhs of Eq.~(\ref{eq:GPD'-D}) is obtained from the rhs of 
   Eq.~(\ref{eq:ffan-D'}) under the replacement 
 $\chi_{\mathbb{PP}}^{\mathbb{P}} \rightarrow 
 \chi_{\mathbb{P}I}^{\mathbb{P}}$.

In the next section, we apply Eqs.~(\ref{eq:sig-2jet-dpe-fact}),
 (\ref{eq:sig-2jet-sd-fact}),
(\ref{eq:sig-2jet-dpe-eik}-\ref{eq:sig-2jet-sd-eik}), and 
(\ref{eq:sig-2jet-dpe}-\ref{eq:sig-2jet-sd}) to investigate the rapidity gap
survival for diffractive dijet production in $pp$ collisions.
We shall use the parameter set of the QGSJET-II-04 model \cite{ost11},
which has been obtained by fitting the model to available accelerator
data on  total and elastic proton-proton cross sections, elastic scattering
slope, and total and diffractive structure functions 
$F_{2}$, $F_{2}^{{\rm D}(3)}$.

\section{Results and discussion\label{sec:results}}

\begin{figure*}
\centering{}
\includegraphics[clip,width=0.95\textwidth,height=6.5cm]{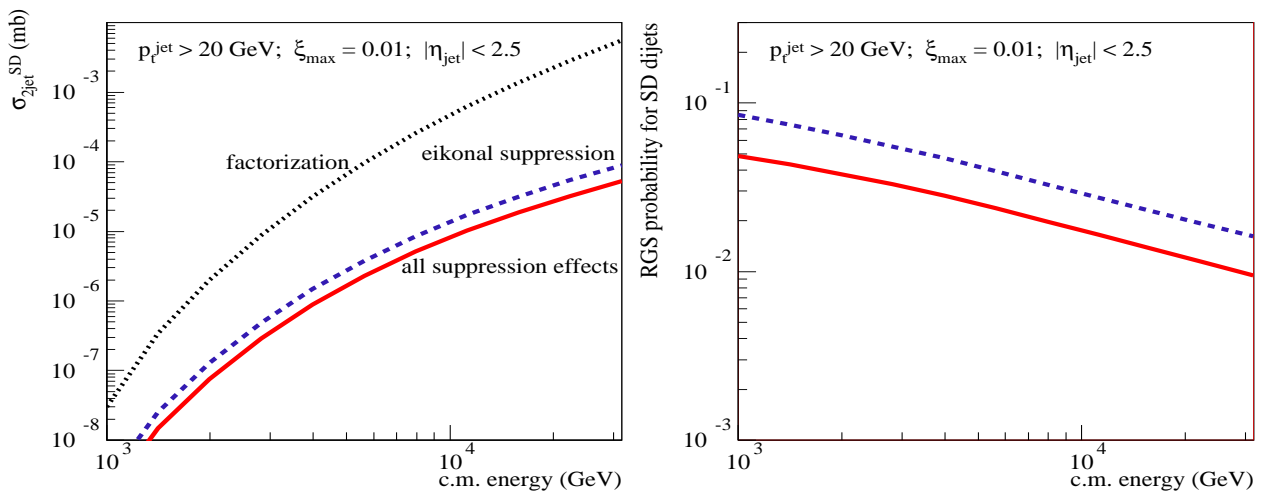}
\caption{Left: energy dependence of the calculated single diffractive 
 dijet cross sections: $\sigma_{pp}^{2{\rm jet-SD(fact)}}$ (dotted),
 $\sigma_{pp}^{2{\rm jet-SD(eik)}}$ (dashed), and 
$\sigma_{pp}^{2{\rm jet-SD}}$ (solid). Right:  energy dependence of the
 corresponding  RGS probabilities:
$S^2_{\rm SD(eik)}$ (dashed) and $S^2_{\rm SD(tot)}$ (solid). All 
for $p_{{\rm t}}^{{\rm cut}}=20$ GeV/c, $\xi_{\max}=0.01$,
 and  $|\eta_{\rm jet}|<2.5$.
\label{fig:sigjet-sd-s}}
\end{figure*}

Let us start with the investigation of the   energy dependence of the  
  dijet production cross section and of the respective rapidity gap survival
probability for single diffractive (SD) proton-proton collisions.
In Fig.~\ref{fig:sigjet-sd-s}~(left), % 
we compare our results for $\sqrt{s}$-dependence of 
$\sigma_{pp}^{2{\rm jet-SD(fact)}}$ 
calculated 
according to Eq.~(\ref{eq:sig-2jet-sd-fact}), based on the factorization
assumption, to the one of $\sigma_{pp}^{2{\rm jet-SD(eik)}}$ 
[Eq.~(\ref{eq:sig-2jet-sd-eik})], which
accounts for the eikonal rapidity gap suppression, and to 
$\sigma_{pp}^{2{\rm jet-SD}}$ 
[Eq.~(\ref{eq:sig-2jet-sd})], which takes into account 
all the above-discussed  suppression effects. We impose here cuts on the
 jet  transverse momentum, 
 $p_{{\rm t}}^{{\rm jet}}>p_{{\rm t}}^{{\rm cut}}=20$ GeV/c,
and on the light cone momentum loss by the projectile proton,
$\xi =M_X^2/s <\xi_{\max}=0.01$, i.e.~$y^{\rm gap}=-\ln \xi_{\max}$,
with $M_X^2$ being the mass squared of the produced diffractive system.
Additionally, we demand both jets to be produced in the central pseudorapidity
$\eta$ region, $|\eta_{\rm jet}|<2.5$.
 In Fig.~\ref{fig:sigjet-sd-s}~(right), we plot the corresponding RGS factors
 $S^2_{\rm SD(eik)}\equiv 
 \sigma_{pp}^{2{\rm jet-SD(eik)}}
 /\sigma_{pp}^{2{\rm jet-SD(fact)}}$
 and  $S^2_{\rm SD(tot)}\equiv
 \sigma_{pp}^{2{\rm jet-SD}} /\sigma_{pp}^{2{\rm jet-SD(fact)}}$.
 While the plotted diffractive dijet cross sections steeply rise with energy,
  due to  the increase of the kinematic space for parton evolution, 
  we observe a  mild energy-dependence for the respective RGS factors.
  Naturally, the probability for the rapidity gap survival goes down at higher
  energies - due to the increase of parton densities, resulting in an
  enhancement of multiple scattering, hence, in a decrease of 
  $S^2_{\rm SD(eik)}$. However, the additional RG suppression by absorptive
  corrections of non-eikonal type, reflected by the ratio
$S^2_{\rm SD(tot)}/S^2_{\rm SD(eik)}\simeq 0.6$, appears to be a much weaker
 and almost energy-independent effect. At the first sight, this seems surprising
 as the energy rise of parton densities should lead to an enhancement of
 rescatterings of intermediate partons from the cascades mediating the 
 diffractive scattering, hence, to stronger  non-eikonal  absorptive
  corrections.

\begin{figure*}
\centering{}
\includegraphics[clip,width=0.95\textwidth,height=6.5cm]{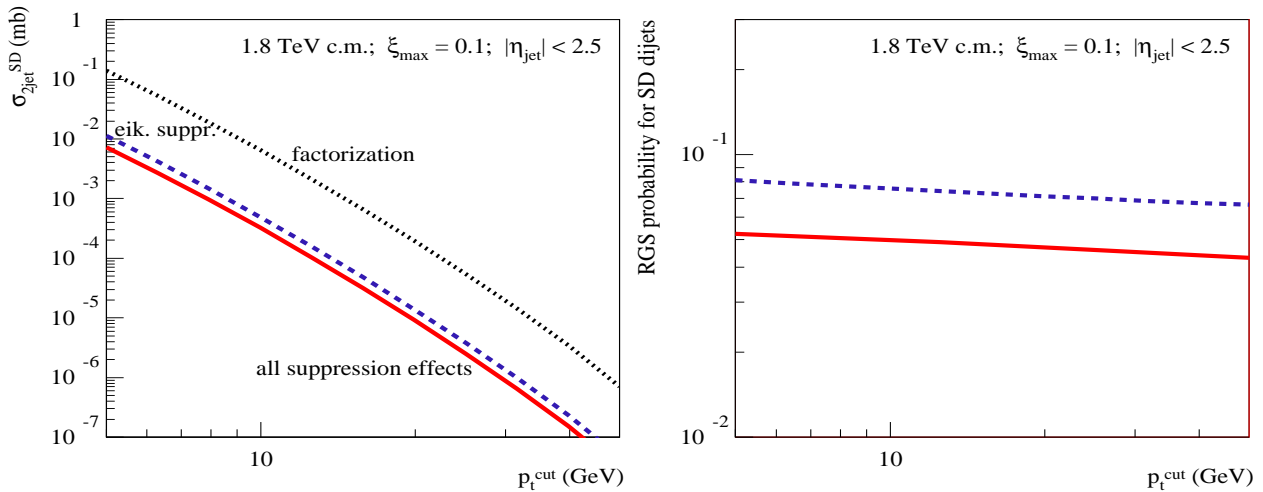}
\caption{Left: $p_{{\rm t}}^{{\rm cut}}$-dependence  of the calculated SD
 dijet cross sections: $\sigma_{pp}^{2{\rm jet-SD(fact)}}$ (dotted),
 $\sigma_{pp}^{2{\rm jet-SD(eik)}}$ (dashed), and 
$\sigma_{pp}^{2{\rm jet-SD}}$ (solid). Right:
 $p_{{\rm t}}^{{\rm cut}}$-dependence of the corresponding RGS probabilities:
$S^2_{\rm SD(eik)}$ (dashed) and $S^2_{\rm SD(tot)}$ (solid). All
for $\sqrt{s}=1.8$ TeV, $\xi_{\max}=0.1$, and  $|\eta_{\rm jet}|<2.5$.
\label{fig:sigjet-sd-pt-tev}}
\end{figure*}

\begin{figure*}
\centering{}
\includegraphics[clip,width=0.95\textwidth,height=6.5cm]{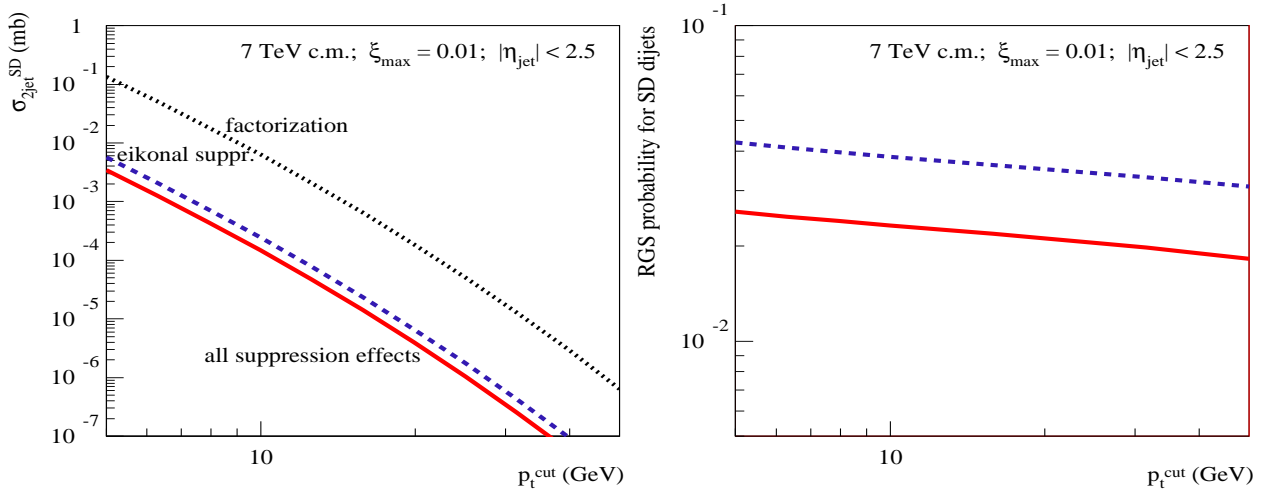}
\caption{Same as in Fig.~\ref{fig:sigjet-sd-pt-tev} for 
$\sqrt{s}=7$ TeV and $\xi_{\max}=0.01$.\label{fig:sigjet-sd-pt}}
\end{figure*}

To get a further insight into the problem, let us check
the dependence of the dijet SD cross sections and of the RGS factors on the
 jet transverse momentum cutoff $p_{{\rm t}}^{{\rm cut}}$ at the energies of
 the Tevatron  (Fig.~\ref{fig:sigjet-sd-pt-tev}), for $\xi_{\max}=0.1$,
   and the LHC (Fig.~\ref{fig:sigjet-sd-pt}),  for $\xi_{\max}=0.01$. % 
The obtained   $p_{{\rm t}}^{{\rm cut}}$-dependencies for both 
$S^2_{\rm SD(eik)}$ and $S^2_{\rm SD(tot)}$ are rather flat. There is a mild
decrease of $S^2_{\rm SD(eik)}$ for increasing  $p_{{\rm t}}^{{\rm cut}}$ -
due to the shift of the dijet production into more opaque region of smaller
impact parameters, which is related to the reduction of the phase space
available for soft ($|q^{2}|<Q_{0}^{2}$) parton evolution, as discussed in
Section~\ref{sec:jet-diffr}. On the other hand, the ratio 
$S^2_{\rm SD(tot)}/S^2_{\rm SD(eik)}$ remains nearly constant over the
studied range 5 GeV/c$<p_{{\rm t}}^{{\rm cut}}<50$ GeV/c. To some extent,
this is less surprising than the flat energy-dependence in 
Fig.~\ref{fig:sigjet-sd-s}~(right), since there are two competing effects here,
both arising from the reduced kinematic phase space for the soft parton
evolution. On the one side, the  shift of the dijet production towards 
 smaller impact parameters should enhance the absorptive effects related
 to rescatterings of intermediate  partons. On the other hand, due to the
reduction of the rapidity space for the soft  parton evolution, one may expect
a weakening of those effects.\footnote{Let us remind that in the current
approach only rescatterings of soft ($|q^{2}|<Q_{0}^{2}$) partons are taken 
into consideration.}

\begin{figure*}
\centering{}
\includegraphics[clip,width=0.95\textwidth,height=6.5cm]{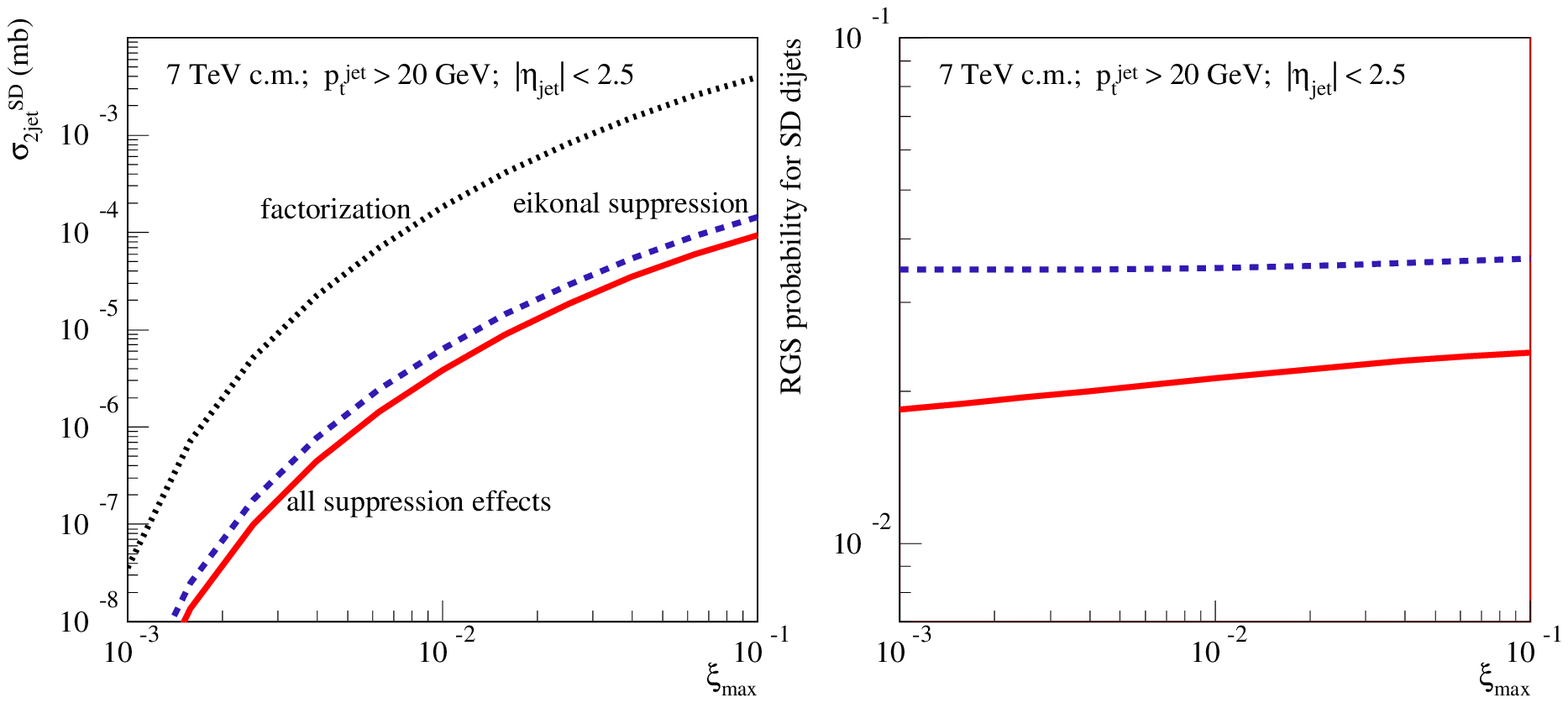}
\caption{Left: $\xi_{\max}$-dependence  of the calculated SD
 dijet cross sections: $\sigma_{pp}^{2{\rm jet-SD(fact)}}$ (dotted),
 $\sigma_{pp}^{2{\rm jet-SD(eik)}}$ (dashed), and 
$\sigma_{pp}^{2{\rm jet-SD}}$ (solid). Right:
 $\xi_{\max}$-dependence of the corresponding RGS probabilities:
$S^2_{\rm SD(eik)}$ (dashed) and $S^2_{\rm SD(tot)}$ (solid). All 
for $\sqrt{s}=7$ TeV, $p_{{\rm t}}^{{\rm cut}}=20$ GeV/c,
 and  $|\eta_{\rm jet}|<2.5$.
\label{fig:sigjet-xgap}}
\end{figure*}

For completeness, let us also study the dependencies of the 
 dijet SD cross sections and of the RGS factors on the size of the 
 rapidity gap. These are plotted in Fig.~\ref{fig:sigjet-xgap} %
 as a function of $\xi_{\max}$, for the production of jets 
 of  $p_{{\rm t}}^{{\rm jet}}> 20$ GeV/c at the LHC energy $\sqrt{s}=7$ TeV.
 Here we observe a rather flat behavior for $S^2_{\rm SD(eik)}$, since 
 the size of the rapidity gap makes a small impact on the slope for the
 diffractive scattering, hence, on the range of impact parameters relevant
 for  SD dijet production. On the other hand,  for decreasing  $\xi_{\max}$ 
 (thus, for an increasing rapidity range
 for virtual parton cascades mediating the diffractive scattering),
 there is some enhancement
 of absorptive effects related to rescatterings of intermediate  partons,
  which results in a slight decrease of the
 RGS probability, with $S^2_{\rm SD(tot)}/S^2_{\rm SD(eik)}$  changing from 
 0.65 for  $\xi_{\max}=0.1$ to 0.53  for  $\xi_{\max}=10^{-3}$.
 
\begin{figure*}
\centering{}
\includegraphics[clip,width=0.95\textwidth,height=6.5cm]{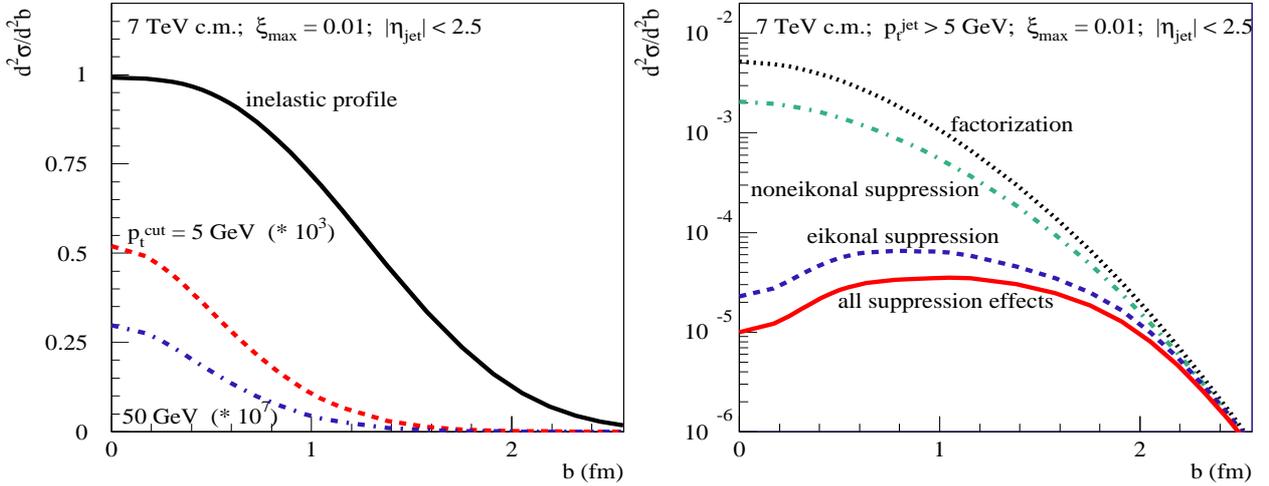}
\caption{Left: transverse profiles for SD dijet production 
$d^2\sigma_{pp}^{2{\rm jet-SD(fact)}}/d^2b$, calculated based on the 
factorization assumption, for 
  $p_{{\rm t}}^{{\rm cut}}=5$ and 50 GeV/c (dashed and dotted-dashed lines,
  respectively),   compared to the inelastic profile (solid line).
  Right:  transverse profiles for SD dijet production for
   $p_{{\rm t}}^{{\rm cut}}=5$ GeV/c, calculated
  taking different absorptive effects into account;
dotted,  dotted-dashed, dashed, and solid lines correspond, respectively, to
$d^2\sigma_{pp}^{2{\rm jet-SD(fact)}}/d^2b$,
$d^2\sigma_{pp}^{2{\rm jet-SD(noneik)}}/d^2b$,
$d^2\sigma_{pp}^{2{\rm jet-SD(eik)}}/d^2b$, and
$d^2\sigma_{pp}^{2{\rm jet-SD}}/d^2b$. All for  $\sqrt{s}=7$ TeV,
 $\xi_{\max}=0.01$, and  $|\eta_{\rm jet}|<2.5$.
\label{fig:rapgap-prof}}
\end{figure*}

 It is clear from our discussion so far that the key to the understanding of
  the rapidity gap suppression of diffractive dijet production is in the
  impact parameter dependence of the respective transverse profiles 
  $d^2\sigma_{pp}^{2{\rm jet-SD(fact)}}/d^2b$,
 $d^2\sigma_{pp}^{2{\rm jet-SD(eik)}}/d^2b$,
 and $d^2\sigma_{pp}^{2{\rm jet-SD}}/d^2b$, which are defined by the
 $b$-integrands of Eqs.~(\ref{eq:sig-2jet-sd-fact}),
(\ref{eq:sig-2jet-sd-eik}), and  (\ref{eq:sig-2jet-sd}), respectively. 
In  Fig.~\ref{fig:rapgap-prof}~(left), %
we plot\\ $d^2\sigma_{pp}^{2{\rm jet-SD(fact)}}/d^2b$ for  $\sqrt{s}=7$ TeV
and  $\xi_{\max}=10^{-2}$, for two values of the jet $p_{{\rm t}}$-cutoff:
 $p_{{\rm t}}^{{\rm cut}}=5$ and 50 GeV/c, 
 in comparison to the inelastic profile
 $G_{pp}^{\rm inel}(s,b)=1-\sum_{i,j}C_i\, C_j\; e^{-\Omega _{ij}(s,b)}$.
 Here we immediately see the origin of the factorization breaking
 for   diffractive dijet production: the respective profile defined by
 the factorization ansatz,  Eq.~(\ref{eq:sig-2jet-sd-fact}), is confined to the
 opaque region of small impact parameter $b$, where the probability of
 additional inelastic rescatterings between the protons' constituents is
 close to unity. When taking into account the eikonal RG suppression,
 i.e.~including the probability for no such rescatterings
 [factor $e^{-\Omega _{ij}(s,b)}$ in the rhs of Eq.~(\ref{eq:sig-2jet-sd-eik})],
  the corresponding production rate at $b\simeq 0$
 is reduced by many orders of magnitude [c.f.~dotted and dashed lines in
  Fig.~\ref{fig:rapgap-prof}~(right)]. At $b> 2$ fm, the absorption 
  becomes weak, yet the production rate is miserable there. Hence, the bulk
  of the SD dijet production comes from the intermediate region
   $b\sim 1-1.5$ fm.
  It is worth stressing that the above-discussed transverse picture is of
  generic character, being a consequence of the fundamental feature of 
  hadronic collisions, namely, that the slope for diffractive scattering
  is considerably smaller than the elastic scattering slope $B_{pp}^{\rm el}$
  which defines the transverse spread of the inelastic profile
   $G_{pp}^{\rm inel}(s,b)$.

Let us next consider the   profile 
$d^2\sigma_{pp}^{2{\rm jet-SD(noneik)}}/d^2b$, plotted as the dotted-dashed line
in Fig.~\ref{fig:rapgap-prof} (right), which corresponds to taking  
 noneikonal absorption into account, while neglecting the eikonal
RG suppression. It is defined by the  $b$-integrand of 
Eq.~(\ref{eq:sig-2jet-sd}), omitting the factors  $e^{-\Omega _{ij}(s,b)}$.
We see that the respective effects, being reflected by the 
differences between the dotted and  dotted-dashed lines in the Figure, are
strongest at small $b$, i.e.~where rescatterings of intermediate partons
are enhanced by a higher parton density. However, as discussed above, the
contribution of the   small $b$ region to the  diffractive dijet production
is strongly suppressed by the eikonal absorption. This explains  the 
relatively  weak effect of the noneikonal absorption, observed in 
Figs.~\ref{fig:sigjet-sd-s}--\ref{fig:sigjet-xgap}. In other words, as
argued in Ref.~\cite{kmr17,kmr09}, the eikonal RG suppression effectively
eliminates the kinematic region where  noneikonal absorptive effects could
be of significant importance. This also helps us to understand the very week
energy-dependence of the noneikonal absorption, observed in 
Fig.~\ref{fig:sigjet-sd-s}~(right). Moving to higher energies,
  diffractive dijet production at relatively small $b$
is stronger and stronger suppressed by  the eikonal RG suppression and
important contributions come only from larger impact parameters where the
 noneikonal absorption becomes weaker.
 
\begin{figure}
\centering{}
\includegraphics[clip,width=0.45\textwidth,height=6.5cm]{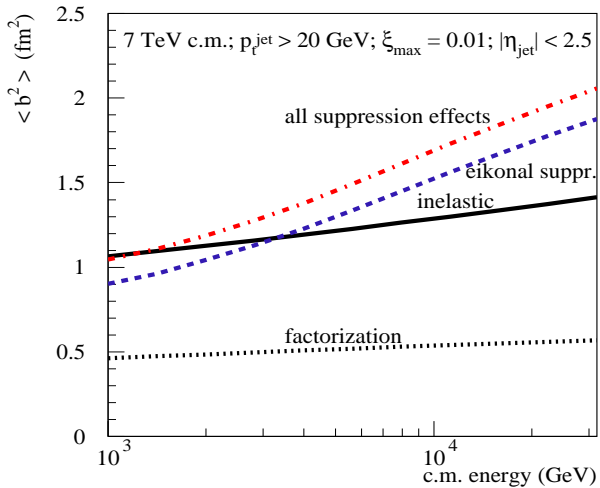}
\caption{Average $b^2$ for inelastic $pp$ collisions (solid) and for
  SD dijet production for $\sqrt{s}=7$ TeV, $p_{{\rm t}}^{{\rm cut}}=20$ GeV/c,
 $\xi_{\max}=0.01$, and  $|\eta_{\rm jet}|<2.5$, as calculated using the 
 different approximations: factorization (dotted), eikonal suppression (dashed),
 and all suppression effects (dotted-dashed).  
\label{fig:rapgap-bb}}
\end{figure}

 The above-discussed tendencies become more clear if we compare the 
 energy-dependence of the average impact parameter squared
 for the different approximations,
  $\langle b_{(X)}^2\rangle=\int \!d^2b\;b^2\,
  \frac{d^2\sigma_{pp}^{(X)}}{d^2b}/\sigma_{pp}^{(X)}$ 
  [$X=$~2jet-SD(fact), 2jet-SD(eik), and 2jet-SD], 
   to the one for general inelastic
  collisions, $\langle b_{\rm inel}^2\rangle=\int \!d^2b\;b^2\, 
  G_{pp}^{\rm inel}(s,b)/\sigma_{pp}^{\rm inel}(s)$,
  as plotted in Fig.~\ref{fig:rapgap-bb}. %
We notice that $\langle b_{\rm (2jet-SD(fact))}^2\rangle$ calculated
based on the factorization assumption is more than twice smaller than
$\langle b_{\rm inel}^2\rangle$. This is not surprising since, firstly,
already the  slope for soft diffraction is considerably smaller than
$B_{pp}^{\rm el}$  and, secondly, in case of hard diffraction a large part
of the available rapidity range is ``eaten'' by hard ($|q^{2}|>Q_{0}^{2}$)
 parton evolution characterized by weak transverse diffusion,
$\Delta b^2 \sim 1/|q^{2}|$,  neglected here.
 Let us also remark that for increasing
 energy $\langle b_{\rm (2jet-SD(fact))}^2\rangle$ rises slower than 
 $\langle b_{\rm inel}^2\rangle$ because the hard parton evolution covers
 a longer and longer rapidity interval. Next, we notice that, firstly,
 $\langle b_{\rm (2jet-SD(eik))}^2\rangle$ is considerably larger than 
 $\langle b_{\rm (2jet-SD(fact))}^2\rangle$ and, secondly, it has a significantly 
 steeper energy rise. This is because the region of small $b$ is strongly
 suppressed by the eikonal absorption
  [c.f.~dotted and dashed lines in Fig.~\ref{fig:rapgap-prof} (right)]
   and for higher energies the strong
 absorption extends towards larger impact parameters, as a consequence
 of the widening and ``blackening'' of the inelastic profile,
 as noticed already in Ref.~\cite{glm98}, which are 
 caused by parton transverse diffusion and the energy rise of parton densities,
 respectively. The same arguments apply to the  noneikonal absorption, due to
 its dependence on parton density: it is  strongest at small $b$ and, 
 for  increasing energy,  becomes more important at larger impact parameters.
   Consequently,  $\langle b_{\rm (2jet-SD)}^2\rangle$ is slightly larger than
  $\langle b_{\rm (2jet-SD(eik))}^2\rangle$.

\begin{figure*}
\centering{}
\includegraphics[clip,width=0.95\textwidth,height=13cm]{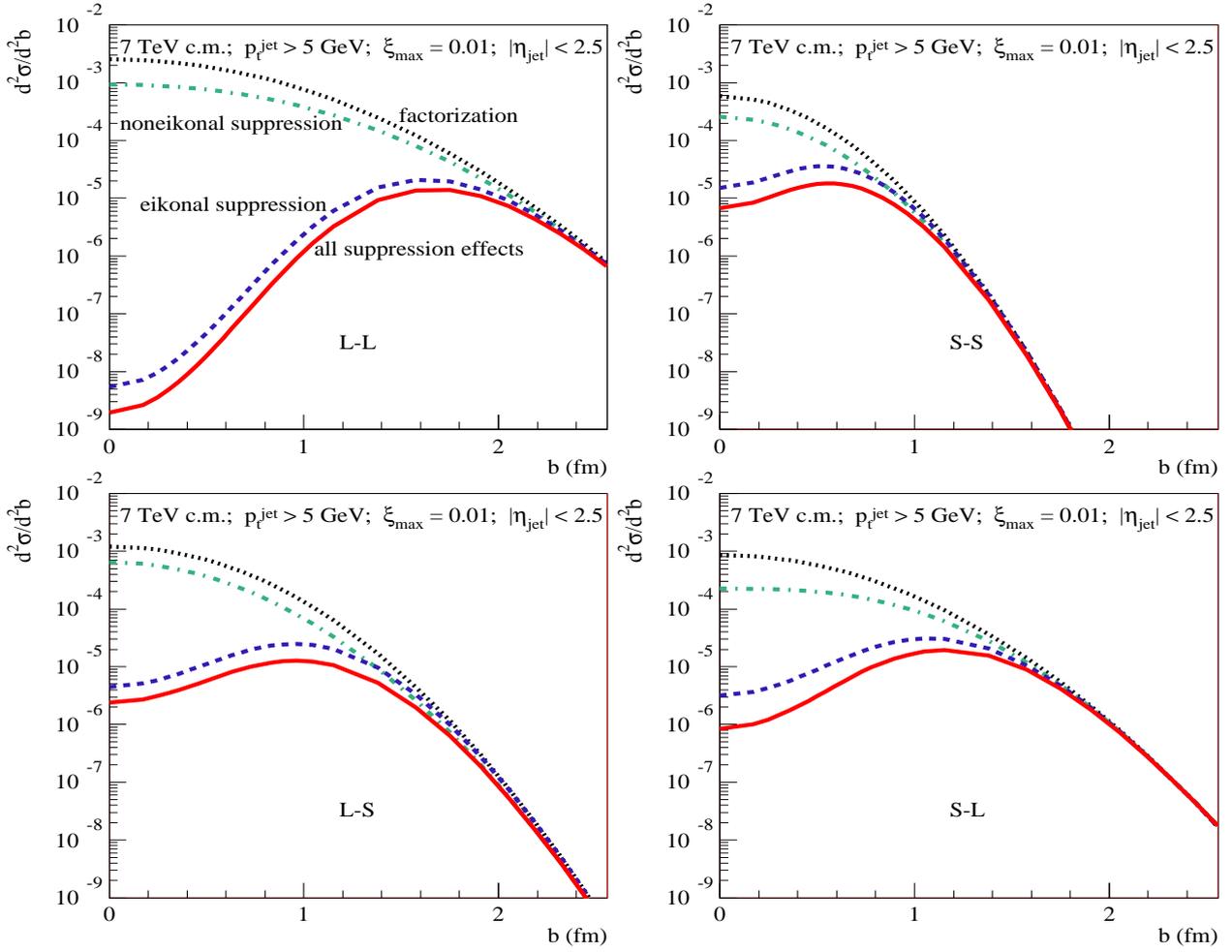}
\caption{Same as in Fig.~\ref{fig:rapgap-prof} (right) for different combinations
of diffractive eigenstates of the projectile and target protons, as explained
in the text.
\label{fig:prof-gw}}
\end{figure*}

As the cross section formulas, Eqs.~(\ref{eq:sig-2jet-sd-fact}),
(\ref{eq:sig-2jet-sd-eik}), and  (\ref{eq:sig-2jet-sd}), are obtained averaging
over contributions of different Fock states of the projectile and  target
protons, it may be  sensible to discuss the relative roles of the
different absorptive  corrections separately for particular combinations
of those  Fock states. This is illustrated in Fig.~\ref{fig:prof-gw}, %
  where we plot  for  $\sqrt{s}=7$ TeV,  $p_{{\rm t}}^{{\rm cut}}=5$ GeV/c,
and  $\xi_{\max}=10^{-2}$ the respective partial contributions  
 $d^2\sigma_{pp(ij)}^{2{\rm jet-SD(fact)}}/d^2b$,
 $d^2\sigma_{pp(ij)}^{2{\rm jet-SD(noneik)}}/d^2b$,\\ 
  $d^2\sigma_{pp(ij)}^{2{\rm jet-SD(eik)}}/d^2b$,
 and $d^2\sigma_{pp(ij)}^{2{\rm jet-SD}}/d^2b$ for the 4 different cases:
  when both the   projectile and the target protons are represented by their
   largest size  Fock states, marked as ``L--L'' in the Figure,
   for an interaction between the small size states (``S--S''),
   and for interactions between  Fock states of different sizes 
    (``L--S'' and ``S--L''). In the ``L--L'' case, we see that all the
    above-discussed tendencies, in particular, the strong suppression of the
    small $b$ region are much more prominent, being enhanced by the larger
    (integrated) parton densities for the large size states.
    On the other hand, in the ``S--S'' case, the interaction profile is
    more transparent due to smaller parton densities, resulting in a weaker
    absorption at small impact parameters.    However, because of
    the smaller scattering slope, the dijet production is confined here to the
  small $b$   region, thus making a small contribution to the overall yield.
  A similar competition between the transverse spread and the strength of
  absorption we observe for the two non-diagonal cases. In the ``L--S'' case,
  the diffractive scattering of the projectile proton is enhanced by its
  larger parton density. Moreover, both the eikonal absorption and the one
  due to intermediate parton rescatterings off the target proton are reduced
  because of the lower parton density for the latter. However, the scattering 
  slope in this case is sizably smaller tahn in the  ``S--L'' case.
 Consequently, the
  latter contribution appears to be a more important one, despite stronger
 noneikonal absorptive corrections related to rescatterings of intermediate
  partons off the target proton which has a higher parton density 
  than in the ``L--S'' case   (c.f.~dotted and dotted-dashed lines in
 the lower right panel of Fig.~\ref{fig:prof-gw}).
  
\begin{figure*}
\centering{}
\includegraphics[clip,width=0.95\textwidth,height=6.5cm]{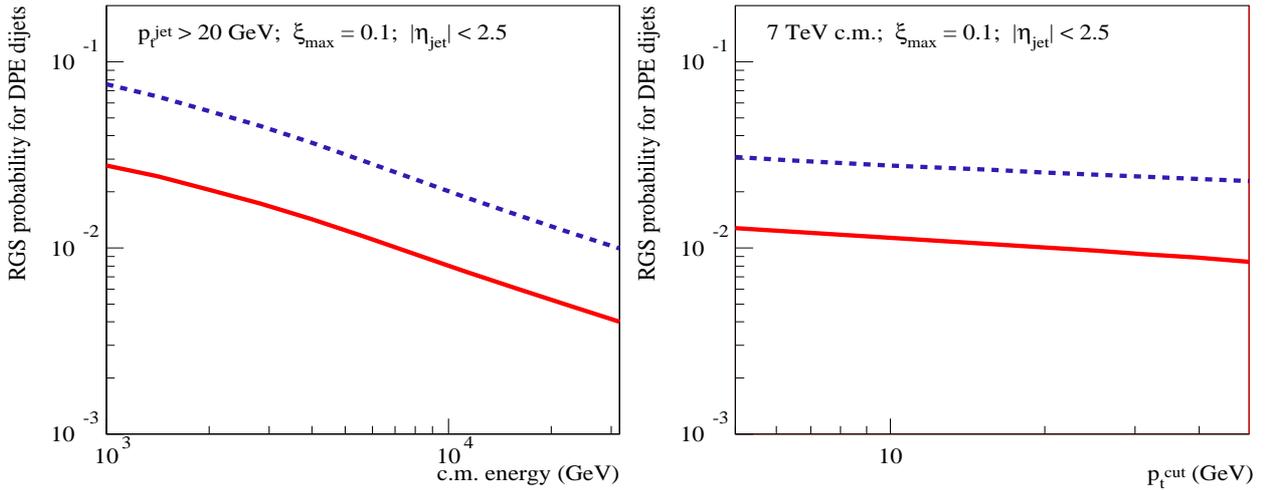}
\caption{Energy  dependence for $p_{{\rm t}}^{{\rm cut}}= 20$ GeV/c (left)
 and $p_{{\rm t}}^{{\rm cut}}$-dependence for $\sqrt{s}=7$ TeV (right)
  of the
RGS factors $S^2_{\rm DPE(eik)}$ (dashed) and $S^2_{\rm DPE(tot)}$ (solid)
for central diffractive dijet production;
 $\xi_{\max}=0.1$,   $|\eta_{\rm jet}|<2.5$.
\label{fig:rapgap-dpe}}
\end{figure*}

\begin{figure}
\centering{}
\includegraphics[clip,width=0.45\textwidth,height=6.5cm]{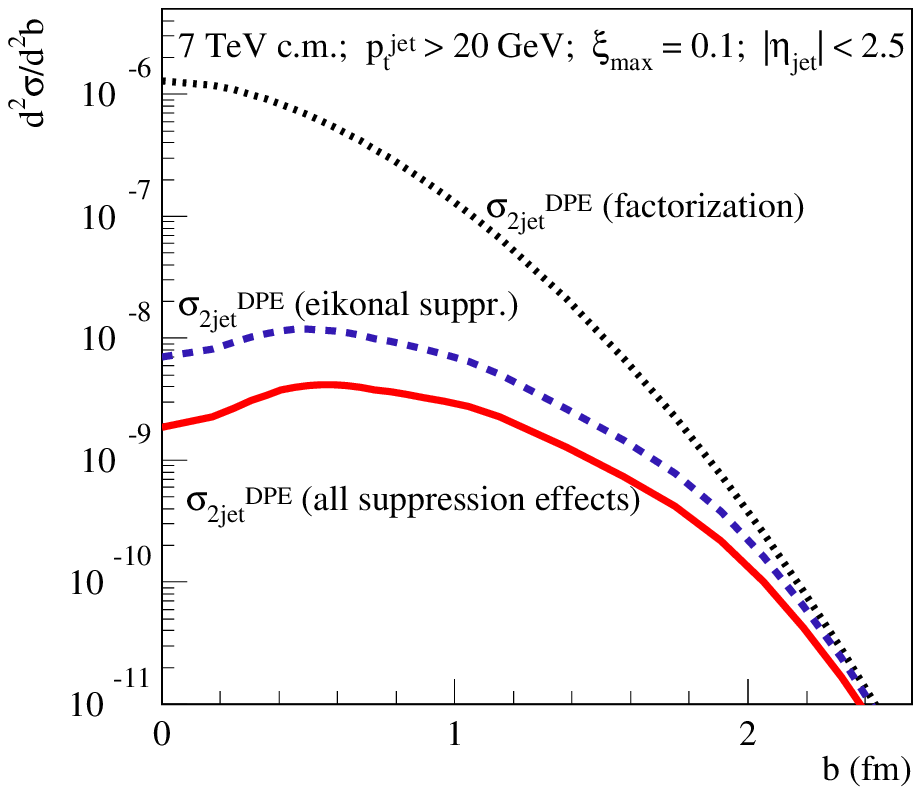}
\caption{Transverse profiles for central diffractive dijet production 
at $\sqrt{s}=7$ TeV for  $p_{{\rm t}}^{{\rm jet}}>20$ GeV/c, calculated
  taking different absorptive effects into account:
$d^2\sigma_{pp}^{2{\rm jet-SD(fact)}}/d^2b$ (dotted),
$d^2\sigma_{pp}^{2{\rm jet-SD(eik)}}/d^2b$ (dashed), and
$d^2\sigma_{pp}^{2{\rm jet-SD}}/d^2b$ (solid);
 $\xi_{\max}=0.1$,   $|\eta_{\rm jet}|<2.5$.
\label{fig:dpe-prof}}
\end{figure}

  Let us now turn to the case of central diffractive dijet production.
In Fig.~\ref{fig:rapgap-dpe}, %  
we plot the energy (for $p_{{\rm t}}^{{\rm jet}}> 20$ GeV/c) and the 
$p_{{\rm t}}^{{\rm cut}}$ (for $\sqrt{s}=7$ TeV) dependencies of the
corresponding  RGS factors (all for $\xi_{\max}=0.1$), 
 $S^2_{\rm DPE(eik)}=  \sigma_{pp}^{2{\rm jet-DPE(eik)}}
 /\sigma_{pp}^{2{\rm jet-DPE(fact)}}$
 and  $S^2_{\rm DPE(tot)}= \sigma_{pp}^{2{\rm jet-DPE}}
 /\sigma_{pp}^{2{\rm jet-DPE(fact)}}$, 
 where\\ $\sigma_{pp}^{2{\rm jet-DPE(fact)}}$, 
 $\sigma_{pp}^{2{\rm jet-DPE(eik)}}$, and $\sigma_{pp}^{2{\rm jet-DPE}}$ 
 are defined by Eqs.~(\ref{eq:sig-2jet-dpe-fact}), 
(\ref{eq:sig-2jet-dpe-eik}), and (\ref{eq:sig-2jet-dpe}), respectively.
Additionally, in Fig.~\ref{fig:dpe-prof} %
we show the corresponding transverse profiles
  $d^2\sigma_{pp}^{2{\rm jet-DPE(fact)}}/d^2b$,
  $d^2\sigma_{pp}^{2{\rm jet-DPE(eik)}}/d^2b$,
 and $d^2\sigma_{pp}^{2{\rm jet-DPE}}/d^2b$, given by the $b$-integrands
 in the rhs of  Eqs.~(\ref{eq:sig-2jet-dpe-fact}), 
(\ref{eq:sig-2jet-dpe-eik}), and (\ref{eq:sig-2jet-dpe}).
Here we observe for the RGS probability the same tendencies as in the case
of single diffraction: a relatively weak dependence on the collision energy
 and a low sensitivity to the jet transverse momentum cutoff, also a nearly
constant ratio $S^2_{\rm DPE(tot)}/S^2_{\rm DPE(eik)}\simeq 0.4$. The overall
absorption is approximately twice stronger, compared to SD dijet production,
 because of the smaller scattering slope for  central diffraction,
 which thus concentrates at smaller impact parameters [c.f.~dotted lines in
   Figs.~\ref{fig:rapgap-prof}~(right) and Fig.~\ref{fig:dpe-prof}].
    There, both eikonal and noneikonal
   absorptive corrections are enhanced by the higher parton densities
   in the projectile and target protons, resulting in a substantial suppression
   of the production profile, as one can see in    Fig.~~\ref{fig:dpe-prof}.
    Interestingly,
   the obtained additional suppression of the RGS probability by noneikonal
   absorptive effects, $S^2_{\rm DPE(eik)}/S^2_{\rm DPE(tot)}\simeq 2.5$,
   fits well in the range [2--3]  estimated earlier in Ref.~\cite{fs07}
   using a different framework.

\begin{figure*}
\centering{}
\includegraphics[clip,width=0.95\textwidth,height=6.5cm]{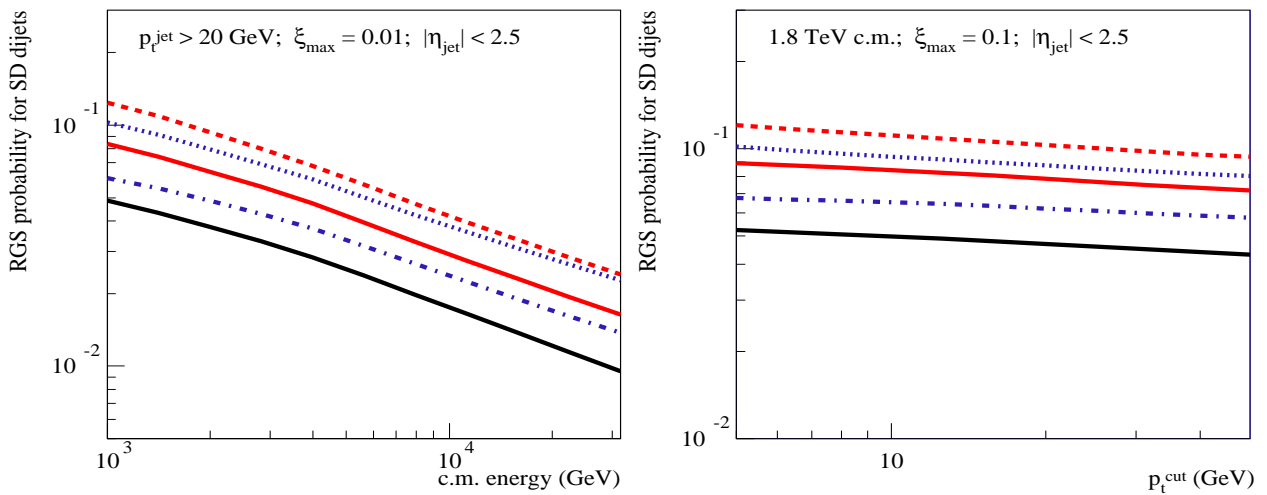}
\caption{Energy dependence for $p_{{\rm t}}^{{\rm cut}}= 20$ GeV/c, 
 $\xi_{\max}=0.01$ (left) 
 and $p_{{\rm t}}^{{\rm cut}}$-dependence for
  $\sqrt{s}=1.8$ TeV, $\xi_{\max}=0.1$
  (right)
  of the RGS factors $S^2_{\rm SD(eik)}$  and $S^2_{\rm SD(tot)}$ 
for SD dijet production, using alternative parameter tunes:
SD- (respectively, dashed and upper solid lines), and SD+  
(respectively, dotted and dotted-dashed lines);
  $|\eta_{\rm jet}|<2.5$. The downmost solid lines correspond to 
   $S^2_{\rm SD(tot)}$ calculated with the default parameters.
 \label{fig:rapgap-opt}}
\end{figure*}

It may be interesting to check how robust are the presented results with
respect to variations of   parameters of the adopted model. To address
that, we repeat the calculations of the RGS probability for SD dijet
production using two alternative parameter tunes of the QGSJET-II-04 model,
discussed in Ref.~\cite{ost14} in relation to present uncertainties of soft
diffraction measurements at the LHC. One of the tunes, referred to as ``SD-'', 
 yields  30\% smaller low mass diffraction cross section, compared to the
 default parameter settings, because of a smaller difference between the
 strengths of the Pomeron coupling to different diffractive eigenstates of the
 proton. At parton level, this would correspond to weaker color fluctuations
 in the proton. The other one, referred to as ``SD+'',
is characterized by an increased rate of high mass diffraction in $pp$ 
collisions, which has been achieved by using a higher value for the
 triple-Pomeron coupling, and a slightly smaller low mass diffraction. 
  Apart from those features, both parameter tunes have been calibrated with the 
 same set of experimental data on hadronic cross sections and particle
 production as the default model (see  Ref.~\cite{ost14} for more details).
  The calculated energy  (for $p_{{\rm t}}^{{\rm cut}}= 20$ GeV/c and
 $\xi_{\max}=0.01$)  and  $p_{{\rm t}}^{{\rm cut}}$ (for $\sqrt{s}=1.8$ TeV
 and $\xi_{\max}=0.1$) dependencies of the corresponding
  RGS factors for SD dijet production  are shown in 
 Fig.~\ref{fig:rapgap-opt}, %
 being very similar to each other and to the above-discussed results
 obtained using the default model parameters. This applies also
 to the relative  importance of the noneikonal absorption: the calculated 
$S^2_{\rm SD(tot)}/S^2_{\rm SD(eik)}$  ranges between 0.6 and 0.8,
depending on the parameter set and the event selection.
 However, the absolute value
 of the RGS probability appears to be quite sensitive to 
  the treatment of low mass diffraction, being  some 70\% higher
 for the SD- tune,   due to a slightly  more transparent inelastic profile,
   compared to the default case.

Let us finally check whether the obtained values for the RGS probability
are compatible with available experimental data. Here the situation is
somewhat confusing. At $\sqrt{s}=1.8$ TeV, using the parameter sets of
QGSJET-II-04, SD+, and SD- tunes, for SD dijet production 
($p_{{\rm t}}^{{\rm cut}}= 7$ GeV/c and  $\xi_{\max}=0.1$) we obtain the values
$S^2_{\rm SD(tot)}\simeq 0.05$, 0.07, and 0.09, respectively 
[see Fig.~\ref{fig:rapgap-opt} (right)], which are
all compatible with the  CDF result, $0.06 \pm 0.02$ \cite{aff00}.
However, recent measurements at the LHC by the CMS \cite{cha13} and
ATLAS \cite{atlas16} experiments indicate that the  RGS probability 
for SD dijet production at $\sqrt{s}=7$ TeV
 ($p_{{\rm t}}^{{\rm cut}}= 20$ GeV/c and
   $\xi_{\max}\sim 10^{-3}$) is at 10\% level, which is compatible with the 
   CDF result at $\sqrt{s}=1.8$ TeV and is almost an order of magnitude higher than 
   what we obtain here [c.f.~Fig.~\ref{fig:rapgap-opt} (left)].
   Thus, we find the experimental situation very puzzling since the decrease
   with energy of the  RGS probability is closely related to the 
   significant shrinkage  of the diffractive cone, convincingly demonstrated
   by the TOTEM \cite{totem17}  and ATLAS \cite{aad14} measurements.
    As the scattering slope for (unabsorbed) diffractive
   dijet production rises with energy slower than  $B_{pp}^{\rm el}$
   (c.f.~dotted and solid lines in Fig.~\ref{fig:rapgap-bb} for the respective
   $\langle b^2 \rangle$),   at higher energies the bulk of the 
   production becomes confined to more and more opaque
   region. If the flat energy behavior  of the  RGS probability
   is further confirmed,   notably, by using also information 
   from proton tagging by forward detectors at the LHC,
   this would imply a very nontrivial dynamics of hadronic collisions.

\section{Conclusions\label{sec:Outlook}}

In this work, we applied the phenomenological Reggeon field theory
framework for calculations of the 
 rapidity gap survival probability for diffractive dijet 
production in proton-proton collisions, investigating in some detail
various absorptive effects contributing to the RG suppression.
Most importantly, we have demonstrated that the absorption due to
  elastic rescatterings of intermediate partons mediating the diffractive
  scattering plays a subdominant role, compared to the eikonal
 rapidity gap suppression due to elastic rescatterings of constituent partons
 of the colliding protons. The corresponding suppression factors,
$S^2_{\rm SD(tot)}/S^2_{\rm SD(eik)}$ and 
$S^2_{\rm DPE(tot)}/S^2_{\rm DPE(eik)}$, are found to depend very weakly
on the collision energy and the event kinematics.
This is good news since, for a given process of interest, one may account
for such effects, in the first crude approximation, via a rescaling
of   jet rates by a constant factor.
 On the other hand,
 such a weak dependence is somewhat accidental, 
as it results from a complex interplay between particular event selections
(e.g.~the choice for the jet $p_{\rm t}$ cutoff) and the corresponding
modification of the  transverse profile for diffractive dijet production.
Generally, such corrections depend on the shape of the transverse profile
for a hard diffraction process of interest and on the kinematic range
available for soft parton evolution, which is influenced, in turn, by the
kinematics of the hard process. For example, we observed 
 $\simeq 40$\% difference for the noneikonal suppression factors
between  the   cases of single and central diffraction. Consequently, 
 our results can not be directly applied to other hard diffraction reactions.

The main suppression mechanism for hard diffraction, related to  
elastic rescatterings of constituent partons of the colliding protons,
is defined by the interplay between the shape of the inelastic profile
for general $pp$ collisions and the transverse profile for a diffractive
process of interest, as already demonstrated previously
 \cite{kmr17,bjo93,kmr01,fs07}. On the other hand, it appears to depend
 sizably on color fluctuations in the proton, which thus introduces a 
 significant model dependence for calculations of the RGS probability.
Reversing the argument, experimental studies of rapidity gap survival
in hard diffraction can provide an insight into the nonperturbative
structure of the proton.

\begin{acknowledgements}
The authors acknowledge useful discussions with V.\ Khoze and M.~Tasevsky. 
This work was supported in part by Deutsche Forschungsgemeinschaft (Project
No.\ OS 481/1-2) and  the State of Hesse via the LOEWE-Center HIC for FAIR.
This work was also partially supported by the COST Action THOR, CA15213.
\end{acknowledgements}

\end{document}